\documentclass{aa}
\usepackage{graphicx}
\usepackage{txfonts}
\usepackage{natbib}

\bibpunct{(}{)}{;}{a}{}{,} 

\begin{document}

\title{X-ray hiccups from Sgr\,A* observed by XMM-Newton}
\subtitle{The second brightest flare and three moderate flares caught in half a day} 
\author{D.\ Porquet\inst{1}
\and N.\ Grosso\inst{1} 
\and P.\ Predehl\inst{2} 
\and G.\ Hasinger\inst{2} 
\and F.\ Yusef-Zadeh\inst{3} 
\and B.\ Aschenbach\inst{2}
\and G.\ Trap\inst{4,5} 
\and F.\ Melia\inst{6} 
\and R.S.\ Warwick\inst{7} 
\and A.\ Goldwurm\inst{4,5} 
\and G.\ B{\'e}langer\inst{8} 
\and Y.\ Tanaka\inst{2} 
\and R.\ Genzel\inst{2}
\and K.\ Dodds-Eden\inst{2}
\and M.\ Sakano\inst{7} 
\and P.\ Ferrando\inst{4,5} 
}

\offprints{D.\ Porquet\ }

\institute{Observatoire astronomique de Strasbourg, Universit{\'e}
              Louis-Pasteur, CNRS, INSU, 11 rue de l'Universit{\'e}, 67000
              Strasbourg, France\\
\email{porquet@astro.u-strasbg.fr}
\and Max-Plank-Institut f{\"u}r extraterrestrische Physik, Postfach
              1312, D-85741, Garching, Germany
\and Department of Physics and Astronomy, Northwestern University, 
             Evanston, IL 60208, USA.
\and CEA, IRFU, Service d'Astrophysique, F-91191 Gif-sur-Yvette, France.
\and Astroparticule et Cosmologie, 10 rue Alice Domont et L{\'e}onie Duquet, F-75205 Paris Cedex 13, France.
\and Department of Physics and Steward Observatory, University of
              Arizona, Tucson, AZ 85721, USA
\and Department of Physics and Astronomy, University of Leicester, Leicester LE1 7RH, UK.
\and XMM-Newton Science Operations Centre, ESA, Villafranca del Castillo, Apartado 78, 28691 Villanueva de la Canada, Spain
}

\date{Received  / Accepted }

\abstract
{Our Galaxy hosts at its dynamical center Sgr\,A*, the
  closest supermassive black hole. 
Surprisingly, its luminosity is
  several orders of magnitude lower than the Eddington
  luminosity. However, the recent observations of occasional rapid X-ray flares
  from Sgr\,A* provide constraints on the accretion and radiation
  mechanisms at work close to its event horizon.
}
{Our aim is to investigate the flaring activity of Sgr\,A* and to
  constrain the physical properties of the X-ray flares.}
{In Spring 2007, we observed Sgr\,A* with XMM-Newton with a 
total exposure of $\sim$230\,ks. We have performed timing and spectral
analysis of the new X-ray flares detected during this campaign.
 To study the range of flare spectral properties, in a consistent manner, 
we have also reprocessed, using the same analysis procedure and the latest calibration,
archived XMM-Newton data of previously reported rapid flares. The 
dust scattering was taken into account during the
    spectral fitting. We also used Chandra archived 
    observations of the quiescent state of Sgr\,A* for comparison.
}
{ On April 4, 2007, we observed for the first time within a time interval of roughly half
  a day, an enhanced incidence rate of X-ray flaring, with a bright flare followed
  by three flares of more moderate amplitude. The former event represents
the second brightest X-ray flare from Sgr A* on record. This new
bright flare exhibits similar light-curve shape (nearly symmetrical),
  duration ($\sim$3\,ks) and spectral characteristics to the very 
  bright flare observed in October 3, 2002 by XMM-Newton. The measured
  spectral parameters of the new bright flare, assuming an absorbed power law model taken into
  account dust scattering effect,  are $N_{\rm
  H}$=12.3$^{+2.1}_{-1.8}\times10^{22}$\,cm$^{-2}$ and
  $\Gamma\sim$2.3$\pm$0.3 calculated at the 90\% confidence level.  
The spectral parameter fits of the sum of the three following moderate 
 flares, while lower ($N_{\rm
 H}$=8.8$^{+4.4}_{-3.2}\times10^{22}$\,cm$^{-2}$ and
 $\Gamma\sim$1.7$^{+0.7}_{-0.6}$), are compatible within the error
 bars with those of the bright flares.  
The column density found, for a power-law model taking into account the dust
 scattering, during the flares is
 at least two times higher than the value 
 expected from the (dust) visual extinction toward Sgr\,A* 
 ($A_{\rm V}\sim25$ mag), i.e., 
 4.5$\times10^{22}$\,cm$^{-2}$. However, our fitting of the Sgr\,A* quiescent
 spectra obtained with Chandra, for a power-law model taking into account the dust
 scattering, shows that an excess of column density is already
 present during the non-flaring phase.  
}
{ The two brightest X-ray flares observed so far from Sgr\,A*
      exhibited similar soft spectra.
}


\maketitle

\section{Introduction}

 Located at the center of our Galaxy, \object{Sgr A*} is the closest
  supermassive black hole to the solar system at a distance of about
  8\,kpc \citep{Reid93,Eisenhauer03,Eisenhauer05}.
Its mass of about 3--4$\times$10$^{6}$\,$M_{\odot}$ has been determined
thanks to the measurements of star motions 
\citep[e.g.,][]{Schoedel02,Ghez03,Ghez05}.
Amazingly, this source is much fainter than expected 
from accretion onto a supermassive black hole.   
Its bolometric luminosity is only
   about 3$\times$10$^{-9}$ L$_{\rm Edd}$ \citep{Melia01,Zhao03}. 
 In particular, its 2--10\,keV X-ray luminosity 
 is  only about 2.4$\times$10$^{33}$\,erg\,s$^{-1}$ 
 within a radius of 1.5$^{\prime\prime}$ \citep{Baganoff03}.
 Thus, Sgr\,A* radiates in X-rays at about 11 orders 
of magnitude less than its corresponding Eddington luminosity. 
  This has motivated the development of various radiatively inefficient
accretion models to explain the dimness of the Galactic Center black hole, 
e.g., Advection-Dominated Accretion Flows 
(e.g., \citealt{Narayan98}), 
jet-disk models (e.g., \citealt{Falcke00}),
 Bondi-Hoyle with inner Keplerian flows (e.g., \citealt{Melia00}). 
 The recent discovery of X-ray flares from Sgr\,A* has provided new 
exciting perspectives for the understanding of the processes at work in the 
Galactic nucleus. 

The first detection of such events was found with
{\sl Chandra} in October 2000. This flare had a duration of  
 about 10\,ks, with a flare peak luminosity of about 1.0 $\times$
 10$^{35}$\,erg\,s$^{-1}$, in the 2--10\,keV energy range,
 i.e., about 45 times the quiescent state \citep{Baganoff01}.   
 Further on several other X-ray flares were detected by {\sl
   XMM-Newton} \citep{Goldwurm03,P03c,Belanger05} 
and {\sl Chandra} \citep{Baganoff03b,Eckart04,Eckart06,Eckart08,Marrone08}.  
The majority of X-ray flares detected up to now have moderate
 flux amplitude with factor of about 10--45 compared to the quiescent
state.
Only one very bright flare with a flux amplitude of about 160 was observed
in October 2002 \citep{P03c}. Remarkably, its peak luminosity of
$\sim$3.6$\times$10$^{35}$\,erg\,s$^{-1}$ was comparable to the
bolometric luminosity of Sgr\,A* during its quiescent state. 
The light curve of the X-ray flares can exhibit short (e.g., 600\,s, 
 \citealt{Baganoff01}; 200\,s, \citealt{P03c}) but deep drops 
close to the flare maximum. This short-time scale could indicate that the
X-ray emission is emitted from a region as small as 7$R_{\rm S}$
 ($\sim$ 13\,$R_{\odot}$).

 In contrast to the near-IR (NIR) flares which appear to be present for
up to 40\% of the time (e.g., \citealt{Genzel03,Ghez04,Yusef06},
Yusef-Zadeh et al.\ 2008 in prep.), the X-ray flaring has a much lower
duty cycle, of typical 1-5\% \citep{Baganoff03b,Belanger05,Eckart06}. 
This means either that the majority of NIR flares have no X-ray
counterpart or that the X-ray-to-NIR ratio of some of the flares is
too small to allow an X-ray detection above the strong, diffuse X-ray
emission from the central parsec.  
When both NIR and X-ray flares are detected simultaneously,  
they show similar morphology in their light curves as well as no apparent delay
between the peaks of flare emission \citep{Eckart04,Eckart06,Eckart08,Yusef06}.
The current interpretation is that both flares come from the same region. 

We report here the results of our Sgr\,A* observation campaign 
performed with {\sl XMM-Newton} from March 30 to April 4, 2007. 
The whole results of this multi-wavelength campaign (VLA, CSO, 
VLT/NACO-VISIR, HST/NICMOS, Integral) will be reported elsewhere 
(Yusef-Zadeh et al.\ 2008 in prep.; Dodds-Eden et al.\ 2008, in prep.).  
We also observed during the three {\sl XMM-Newton} observations, two 
 bright transient sources in outburst \citep{P07b}.
The source located at about 90$^{\prime\prime}$--SW from Sgr\,A*
 has been  associated with the eclipsing X-ray burster \object{AX
   J1745.6-2901}. Seven deep eclipses were observed as well as  
 type-I bursts. The second source is located at about 10$^{\prime}$--NW
from Sgr\,A* and has been associated with the neutron star low-mass
X-ray binary \object{GRS\,1741.9-2853} (a.k.a.\ \object{AX
  J1745.0-2855}). The data analysis of these two sources will be reported in
 forthcoming papers. 

We report, here, for the first time a high level of flaring activity with
 four X-ray flares, one bright and three moderate, detected in 
 half a day. The bright flare is the second brightest X-ray flare detected so
 far from Sgr\,A*.

 In $\S$\ref{sec:data} we describe the observations and data reduction
 procedure used in this work. In $\S$\ref{sec:results} we report the
 timing analysis of Sgr\,A*, and the spectral analysis of the bright
 flare and the sum of the three following moderate flares
 observed during this Spring 2007 campaign. 
 In $\S$\ref{sec:comparison}, we perform a homogeneous and
 self-consistent comparison of spectral properties of the
 X-ray flares of this campaign with the reprocessed data of
   X-ray flares previously observed with {\sl XMM-Newton}. 
 Finally, in $\S$\ref{sec:discussion} we summarize our main
 results and discuss their possible implications.   
 
\section{XMM-Newton observations and data reduction}\label{sec:data}

We observed Sgr\,A* three times with {\sl XMM-Newton} in Spring 2007 for
a total exposure of $\sim$230\,ks. The journal of the {\sl XMM-Newton}
observations is given in Table~\ref{tab:log}.
The EPIC cameras, 2 MOS \citep{Turner01} and one pn \citep{Struder01},
were operated in the full frame window mode with the medium filter.
We use the version 7.1 of the Science Analysis Software (SAS) package
for the data reduction and analysis, with the latest release of the Current
Calibration files (CCF). The MOS and pn event lists were produced
using the SAS tasks {\tt emchain} and {\tt epchain}.
The detector light curves in the 7--15\,keV energy range computed by
these tasks show that the levels of background proton flares were high
only during the last seven hours of the second and the third
observations, where the count rate exceeded the detector telemetry
limit, triggering the counting mode.

For the timing analysis, the contribution of the background proton
flares was estimated using a $\sim$$3\arcmin\times3\arcmin$
area with a low level of X-ray extended-emission, located at
$\sim4\arcmin$-North of Sgr\,A* on the same CCD, where the X-ray
  emission of point sources were subtracted.

\begin{table}[!Ht]
\caption{{\sl XMM-Newton} observation log for the Spring 2007 campaign.}
\label{tab:log}
\vspace{-0.7cm}
\begin{center}
\begin{tabular}{@{}c@{\ }@{\ }c@{\ }@{\ }l@{\ }@{\ }l@{\ }c}
\hline
\hline
\noalign {\smallskip}                       
 {\small Orbit} & \multicolumn{1}{c}{\small ObsID }    &  \multicolumn{1}{c}{\small Start time }
&\multicolumn{1}{c}{\small End time}&\multicolumn{1}{c}{\small duration} \\
 &      &  \multicolumn{1}{c}{\small (UT) }
  &\multicolumn{1}{c}{\small (UT)}&\multicolumn{1}{c}{\small (ks)} \\
\noalign {\smallskip} 
\hline
\noalign {\smallskip} 
{\small 1338} &  {\small 402430701} & {\small Mar 30, 21:29:18.1}  & {\small Mar 31, 06:28:20.8} &  {\small 32.3} \\
{\small 1339} &  {\small 402430301} & {\small Apr 1, 15:09:03.5}  & {\small Apr 2, 19:54:45.8} & {\small 103.5} \\
{\small 1340} &  {\small 402430401} & {\small Apr 3, 16:40:21.5}  & {\small Apr 4, 19:47:39.2} &  {\small 97.6} \\
\noalign {\smallskip} 
\hline
\end{tabular}
\end{center}
\end{table}

\begin{figure*}[!Ht]
\centering
\includegraphics[trim=0 0 12 0,clip,angle=90,width=2\columnwidth]{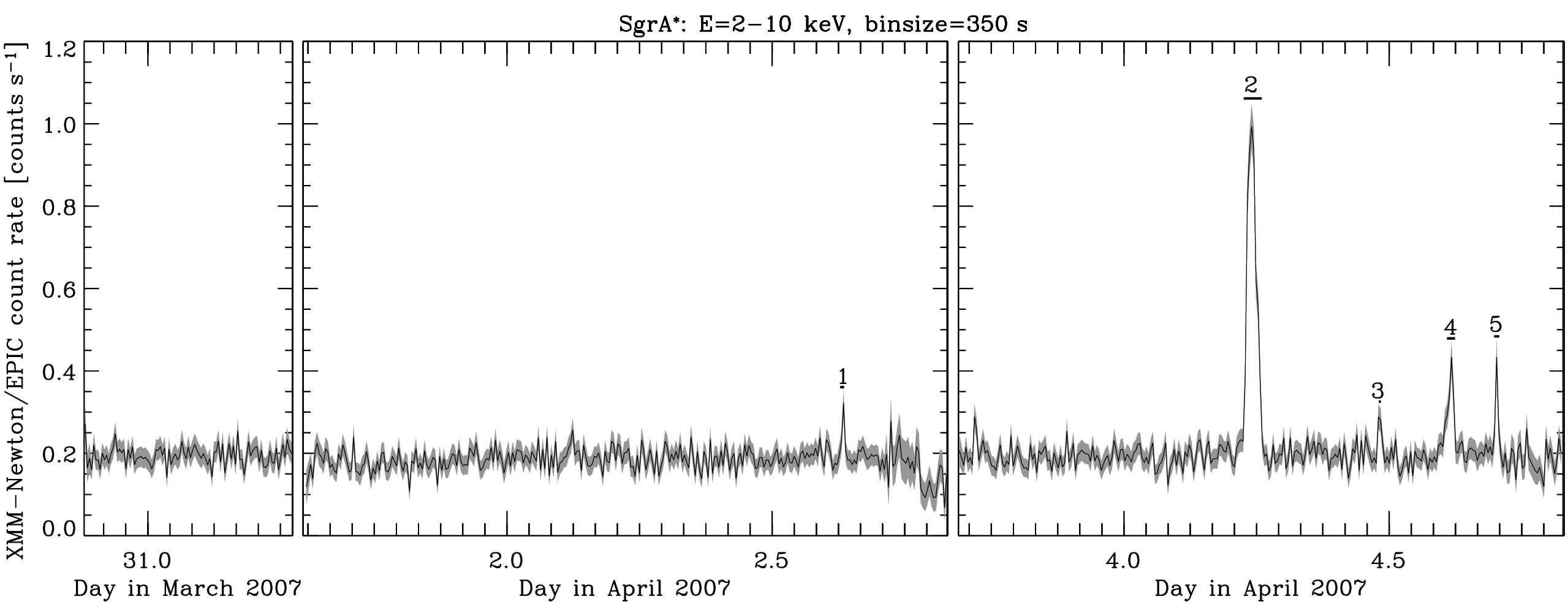}
\caption{{\sl XMM-Newton}/EPIC (pn+MOS1+MOS2) light
  curves of Sgr\,A* in the 2--10\,keV energy range obtained in Spring
  2007. The light curves are corrected from soft-proton flaring
  background. The time interval
  used to bin the light curve is 350\,s. The X-ray flares are labeled
  from 1 to 5. The horizontal lines below these labels indicate the
  flare durations. The quiescent level of Sgr\,A* corresponds to
  only 10\% of the non-flaring level of these light curves (see text for details).}. 
\label{fig:epic_lc}
\end{figure*}

To optimize in our three observations the selection of the events
from Sgr\,A*, we used a two-step method based on the flaring activity of
Sgr\,A* and the presence of the two bright transient sources. We first
computed for each observation a sky image in the 2--10\,keV energy
range that was used to define a circular region of 10$\arcsec$-radius
centered on the X-ray source associated with Sgr\,A*. From this
extraction region, we built a preliminary background-subtracted light
curve\footnote{We selected for MOS and pn the events with 
{\tt PATTERN$\leq$12 and \#XMMEA\_SM}, and {\tt PATTERN$\leq$4 and
  FLAG=0}, respectively.} to identify any flare from Sgr\,A*. We
identified a bright flare from this extraction region in the last observation. 
Its position in each detector was determined 
with the SAS task {\tt edetect\_chain} using only the flare time
interval, while the positions of the other X-ray sources were determined
using the whole exposure of the last observation.
The angular distances between the flaring
X-ray source associated with Sgr\,A* and the two bright transient sources
(e.g., for pn $87.5\arcsec\pm0.2\arcsec$ and $599.9\arcsec\pm0.2\arcsec$,
respectively) were then used to determine the relative position of
Sgr\,A* in the two other observations. This method is independent to
any variation of the detector position angle. 
 Then, we use the X-ray counterparts of the Tycho-2 catalog's sources
\citep{hog00} to obtain an absolute astrometry (see Appendix~\ref{app:astrometry} for
details). The absolute position of the X-ray bright flare is $\alpha_{\rm
J2000}= 17^{\rm h}45^{\rm m}40.0^{\rm s}$, $\delta_{\rm J2000}=
-29^{\circ}00^{\prime}28.6\arcsec$ with a one-sigma
positional uncertainty of $1\farcs0$, i.e., only $0\farcs5$ from
the radio position of Sgr\,A* \citep{Yusef99}. Therefore, the
position of the X-ray bright flare is fully consistent with the radio
position of Sgr\,A*.

\section{Results}\label{sec:results}

\subsection{X-ray light curves of Sgr\,A*}\label{sec:lc}

For each observation and detector, we first built the
source+background and the background light curves in the 2--10\,keV energy range
with 1\,s time bins starting (stopping) at the first (last) good time
interval (GTI) of the corresponding CCD. 
We rebinned the light curves to 350\,s to increase the S/N. Then, we
subtracted the background light curve (scaled to the 
same source extraction area) from the source+background light curve.
We scaled up, using GTI information, count rates and errors affected
by the lost of exposure (e.g., due to the switch from science to
counting mode).

Finally, the background-subtracted light curves of the three detectors
were summed to produce the EPIC light curves. Any detector missing
value was inferred by the one observed by the other detectors using a
(median) scaling factor between the detectors.

The EPIC (pn+MOS1+MOS2) background subtracted light curves of Sgr\,A*
in the 2--10\,keV energy range, with a time bin interval of
  350\,s, are shown in Fig.~\ref{fig:epic_lc}.
During the first half of the exposure, the light curve is almost flat,
with a non-flaring level of X-ray emission consistent with the
one observed with {\sl XMM-Newton} in 2000 \citep{Goldwurm03} and
  2002 \citep{P03c}. The $\sim$50\% higher non-flaring level observed
 in 2004 with {\sl XMM-Newton} \citep{Belanger05}
was due to the contamination by a close transient source in
  outburst \citep{Muno05, P05b}, see also
$\S$\ref{sec:comparison}. Excluding Sgr\,A* and any transient sources
in outburst, the X-ray emission inside the $10\arcsec$-radius region
centered on Sgr\,A* comes mainly from one point source associated with
the complex of stars \object{IRS\,13}, the candidate pulsar wind nebula
\object{G359.95-0.04}, and a diffuse component 
\citep{Baganoff03,Muno03,Wang06}, which contribute to about 90\% of
the non-flaring level in the 2--10\,keV energy range.

\begin{figure}[!Ht]
\centering
\includegraphics[width=0.75\columnwidth]{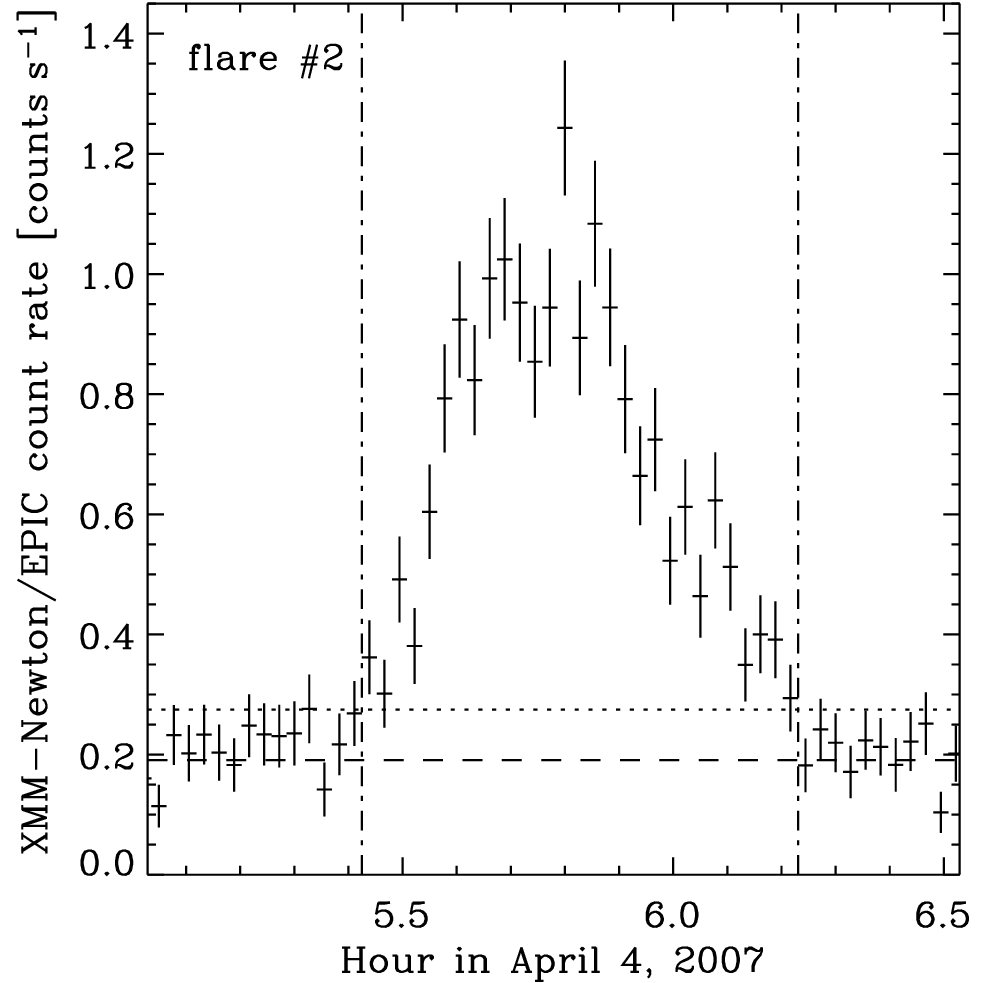}
\caption{{\sl  XMM-Newton}/EPIC light curve of the bright flare in the
  2--10\,keV energy range. The light curve is corrected from
  soft-proton flaring background. The bin time
  interval is 100\,s. The dashed and dotted lines indicate the mean
  non-flaring level and the 1.64$\sigma$ threshold, respectively
  (see text for details). The
  dotted-dashed vertical lines indicate the time interval where the
  light curve deviates from its mean level at a confidence level of
  95$\%$.}
\label{fig:zoom_flare2}
\end{figure}

\begin{figure}[!Ht]
\centering
\begin{tabular}{@{}cc@{}}
\includegraphics[width=0.48\columnwidth]{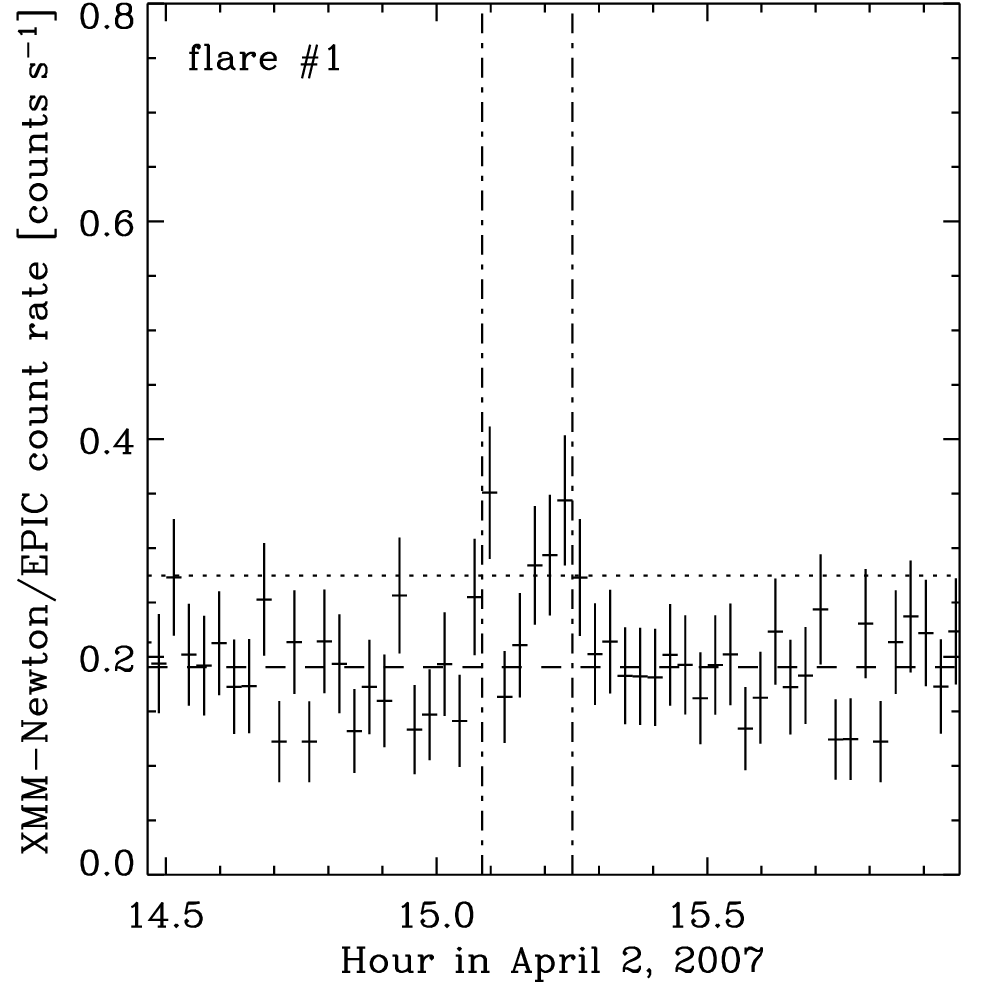}
& \includegraphics[width=0.48\columnwidth]{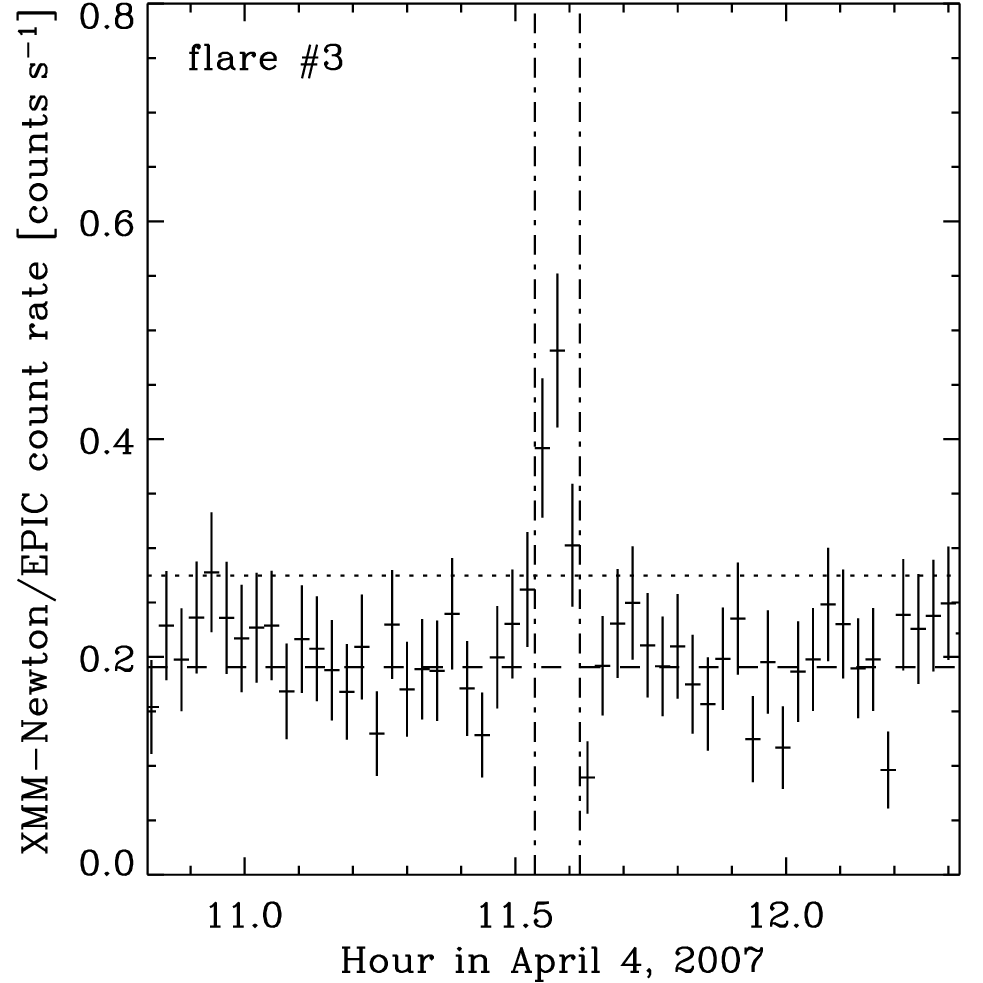} \\
\includegraphics[width=0.48\columnwidth]{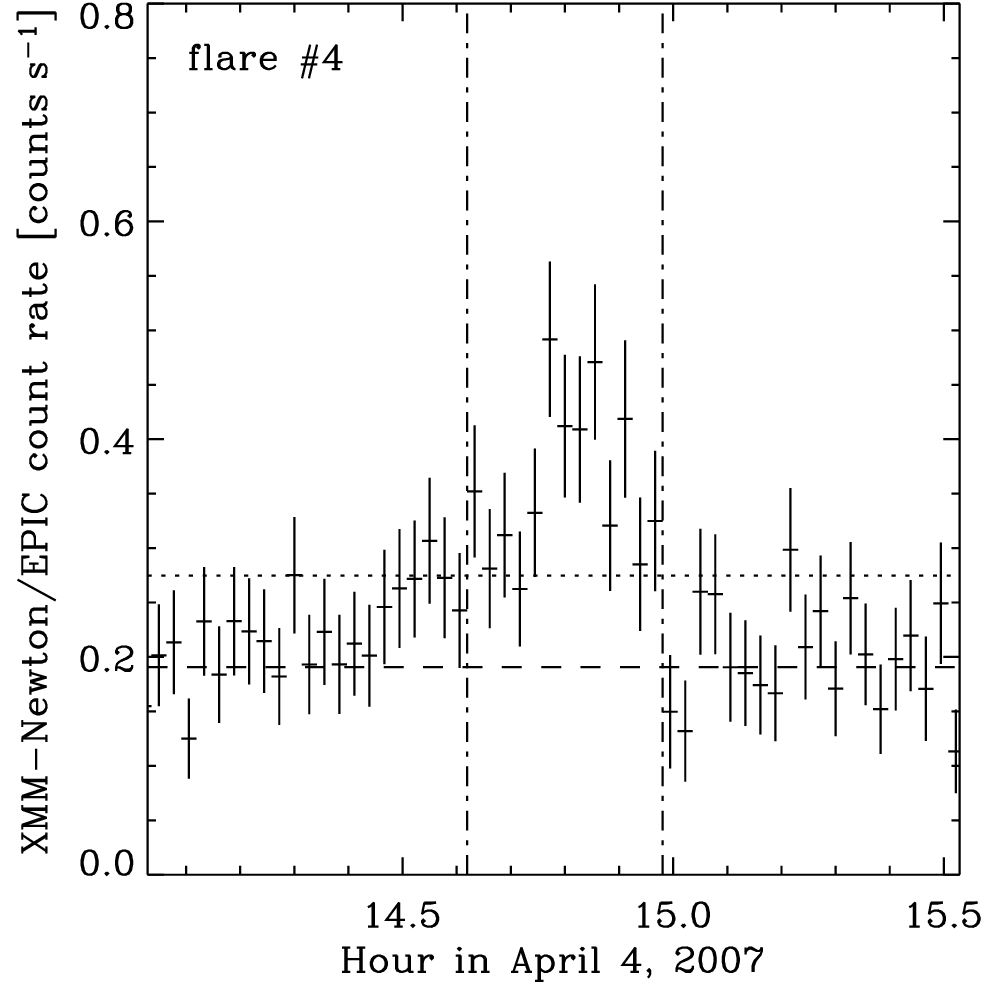}
& \includegraphics[width=0.48\columnwidth]{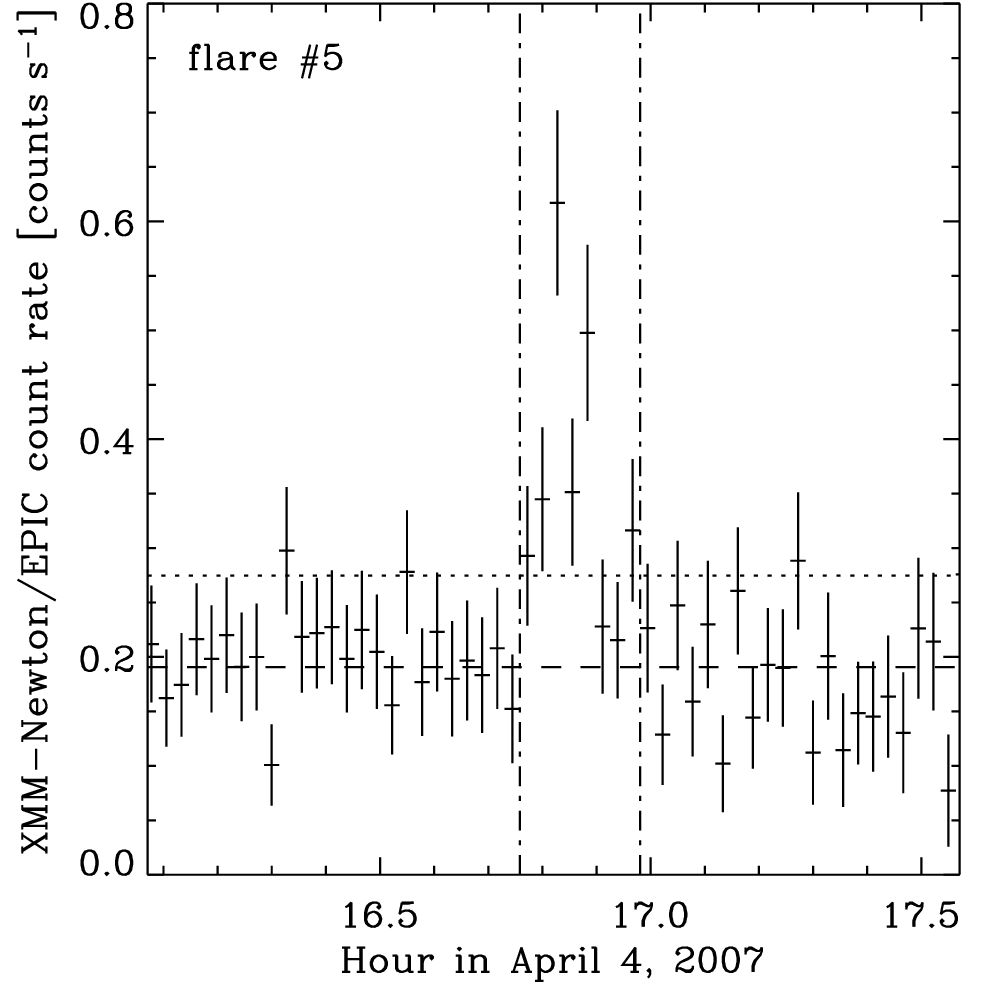}
\end{tabular}
\caption{{\sl XMM-Newton}/EPIC background-subtracted light
  curve of the moderate flares in the 2--10\,keV energy range. Same
  definitions, and time duration, as in Fig.~\ref{fig:zoom_flare2}.}
\label{fig:zoom_flare}
\end{figure}

 We identify any significant deviation from the non-flaring level
  as possible flare. A possible ($\#1$) was observed on April 2, 2007. 
On April 4, one bright flare ($\#2$) was observed followed by three moderate
flares ($\#3$, $\#4$, and $\#5$).
Figures~\ref{fig:zoom_flare2} and \ref{fig:zoom_flare} focus on
the flare light curves with a bin time interval of 100\,s  that we use
  to characterize these flares. We use an iterative 
sigma-clipping of the light curve of each observation to compute the
mean non-flaring level (dashed line) and its standard deviation.
We define the flare time interval as the period where the light curve
deviates from the mean non-flaring level at a confidence level
of 95\% (i.e., where the light curve is above its mean non-flaring
  level plus 1.64 times its standard 
deviation). Table~\ref{tab:flaretime} gives the characteristics of these
X-ray flares. The weak flare $\#1$ peaks just above 3 sigma, and
  it is, therefore, considered as reliable. This weak flare was independently
  confirmed by a simultaneous detection in IR by HST/NICMOS
   (Yusef-Zadeh at al. 2008, in preparation).

The bright flare has a duration of about 3\,ks, similar to the
duration that was observed for the (brightest) flare of October 2002
\citep{P03c}. Its light curve is almost symmetrical, but no
significant deep drop (i.e., about 50\% of flux decrease) is
observed in contrast to the flares of September 2000 
\citep{Baganoff03} and October 2002 \citep{P03c}. 
 Indeed, in the case of the October 2002 flare, the significant drop
was observed in all three instrument light curves, while the apparent drop
in the EPIC light curves of flare $\#$2 is only
observed in the pn light curve. However, for the flare $\#$2 we
cannot rule out a moderate drop in the X-ray light curve.
We also computed the hardness ratio using the 2--5\,keV and 5--10\,keV 
energy ranges, but we found no significant spectral change during the flare interval.

The peak count rates of the moderate flares are 2--6 smaller than that
 of the bright flare.
The durations of the moderate flares are 2--10 times shorter than of
the bright one. The time gaps between two consecutive
flares starting from flare $\#$2 are 5.3, 3.0 and 1.8
  hours. Therefore, four flares were observed in a time interval of only half a day. 
A similar group of three moderate flares were already observed with
  {\sl XMM-Newton} on 2004 March 31 \citep{Belanger05}, but no such
  preceding bright flare was observed. This is the first time that a
  such level of X-ray flaring activity from Sgr\,A*, both in amplitude and
  frequency, is reported.

 When the time coverage of the NIR observations allowed
 simultaneous observations with {\sl XMM-Newton}, the NIR counterparts of
 these X-ray flares have been observed: flares $\#1$, $\#4$, and $\#5$
 were observed with HST/NICMOS (Yusef-Zadeh et al.\ 2008, in prep.);
 and flare $\#2$ was observed with VLT/NACO (Dodds-Eden et al.\ 2008,
 in prep.).

\subsection{Spectral analysis of the X-ray flares}\label{sec:spectral}

\begin{table*}[!Ht]
\caption{Characteristics of the X-ray flares observed in April 2007
  (see Fig.~~\ref{fig:zoom_flare2} and \ref{fig:zoom_flare}). 
}
\label{tab:flaretime}
\begin{center}
\begin{tabular}{@{}c@{}c@{}c@{}cccrrrrcc@{}}
\hline
\hline
\noalign {\smallskip}                       
\multicolumn{1}{c}{\small Flare} 
&  \multicolumn{1}{c}{\small Day}
&\multicolumn{2}{c}{\small Start time$^\mathrm{(a)}$}
&\multicolumn{2}{c}{\small End time$^\mathrm{(a)}$}
&\multicolumn{1}{c}{\small Duration}
&\multicolumn{1}{c}{\small Total$^\mathrm{(b)}$}
&\multicolumn{1}{c}{\small Peak$^\mathrm{(c)}$} 
&\multicolumn{1}{c}{\small Det.$^\mathrm{(d)}$}
&\multicolumn{1}{c}{\small $L_{\rm 2-10~keV}^{\rm peak~(e)}$}
&\multicolumn{1}{c}{\small Ampl.$^\mathrm{(f)}$}\\
\noalign {\smallskip}                       
\multicolumn{1}{c}{\small \#} 
&  \multicolumn{1}{c}{\small }
&\multicolumn{1}{c}{\small  hh:mm:ss}
& \multicolumn{1}{c}{\small s} 
&\multicolumn{1}{c}{\small hh:mm:ss}
&\multicolumn{1}{c}{\small s} 
& \multicolumn{1}{c}{\small s}
&\multicolumn{1}{c}{\small cts}
&\multicolumn{1}{c}{\small cts\,s$^{-1}$}
&\multicolumn{1}{c}{\small $\sigma$}
&\multicolumn{1}{c}{\small 10$^{34}$\,erg\,s$^{-1}$}\\
\noalign {\smallskip} 
\hline
\noalign {\smallskip} 
{\small 1}& {\small 2} & {\small 15:05:03}&291913503  &
{\small 15:15:03}& 291914103 & {\small 600}  & {\small 50.9} & {\small
  0.160}& {\small 3.1} & 3.3$^{+1.2}_{-0.7}$ & 14$^{+5}_{-3}$\\     
\noalign {\smallskip}                       
{\small 2} & {\small 4} & {\small 05:25:30}&292051530   &
{\small 06:13:50}& 292054430  & {\small 2900} & {\small 1443.0} & {\small 1.052}& {\small 20.6} & 24.6$^{+4.8}_{-3.3}$ & 103$^{+20}_{-14}$\\     
\noalign {\smallskip}                       
{\small 3} & {\small 4} & {\small 11:32:10}& 292073530  &     {\small 11:37:10}& 292073830 & {\small 300}  & {\small 60.4} & {\small 0.291}& {\small 5.7} & 6.1$^{+2.2}_{-1.2}$ & 25$^{+9}_{-5}$ \\     
\noalign {\smallskip}                       
{\small 4} & {\small 4} & {\small 14:37:10}&292084630   &     {\small 14:58:50}&292085930 & {\small 1300} & {\small 212.1} & {\small 0.301}& {\small 5.9} & 6.3$^{+2.3}_{-1.3}$ & 26$^{+9}_{-5}$\\     
\noalign {\smallskip}                       
{\small 5} & {\small 4} & {\small 16:45:30}& 292092330  &     {\small 16:58:50}& 292093130 & {\small 800}  & {\small 127.6} & {\small 0.426}& {\small 8.4} & 8.9$^{+3.2}_{-1.8}$ & 37$^{+13}_{-7}$\\  
\noalign {\smallskip} 
\hline
\end{tabular}
\end{center}
\vspace*{-0.5cm}
\begin{list}{}{}
\item[$^{\mathrm{(a)}}$] Start and end times of the flare time
  interval defined as the period where the 
  EPIC light curve deviates from the non-flaring level at a confidence of
  95\%. See text for details.  
\item[$^{\mathrm{(b)}}$] Total EPIC counts in the
  2--10\,keV energy band obtained during the flare interval
   after subtraction of the non-flaring level.
\item[$^{\mathrm{(c)}}$] EPIC count rate in the 2--10\,keV energy band
  at the flare peak after subtraction of the non-flaring level. 
\item[$^{\mathrm{(d)}}$] Detection level at the flare peak in $\sigma$.
\item[$^{\mathrm{(e)}}$] 2--10\,keV luminosity at the flare peak,
  assuming an absorbed power law model (taking account dust scatter
  effect), see parameter fits in Table~\ref{tab:fitCstat}. 
\item[$^{\mathrm{(f)}}$] Amplitude of the flare defined as the ratio of
the 2--10\,keV flare luminosity at the peak and the 2--10\,keV
quiescent luminosity of Sgr\,A* observed with {\sl Chandra} 
(i.e., 2.4 $\times$ 10$^{33}$ erg\,s$^{-1}$, \citealt{Baganoff03}, and
this work, see $\S$\ref{sec:comparison}). 
\end{list}
\end{table*}

We report here the spectral analysis of the four flares observed
during the third observation. We used as extraction region for each
instrument, a 10$^{\prime\prime}$-radius region centered on the
  position of Sgr\,A* determined during the bright flare time
interval ($\S$\ref{sec:data}). The determination of the time
interval of the X-ray flare is crucial to prevent from any bias in the
spectral analysis. For example, in case the flare time interval includes a
significant fraction of non-flaring level, we find that 
both column density and power-law slope values are bias.
This is more important for moderate and weak flares. 
We selected X-ray events with FLAG equal to zero, and since pile-up
was negligible, even during the bright flare, with patterns 0--12 and
0--4 (single and double) for the MOS and pn, respectively.
 To extract the background spectrum we used the same region but
  limited to the non-flaring time interval.
The response matrices and ancillary files were computed using the SAS tasks 
{\tt rmfgen} and {\tt arfgen}, respectively. 
 We used {\tt XSPEC} (version 12.4.0;
\citealt{Arnaud96}) to fit the spectrum with X-ray emission models. 
We report in Table~\ref{tab:fitCstat} the spectral analysis of
the bright flare $\#2$, and of the sum of the three following moderate
flares (i.e., $\#3$+$\#4$+$\#5$) to increase the statistics. 
Instead of using the $\chi^{2}$ statistic, which is not
  appropriate for the fitting of spectra with low counts (e.g., $\#3+\#4+\#5$), we use 
a modified version of {\tt cstat} statistic \citep{Cash79} called
the $W$ statistic \citep{Wachter79} that is implemented in XSPEC 
(Arnaud, in prep.)\footnote{Draft available at\\ 
{\tt ftp://lheaftp.gsfc.nasa.gov/pub/kaa/stat\_paper.ps}\,.}. 
The $W$ statistic is valid for
{\it unbinned} background-subtracted spectra. Binning would erase
information and bias the fitting result.  
We show in online Appendix \ref{app:stat} that
for rather bright flares (e.g., flare $\#2$) 
the parameter fits obtained using this $W$
statistic are very similar to that found using the $\chi^{2}$
statistic (Table~\ref{tab:fit}); but the error bars obtained 
using the $W$ statistic are much better constrained.  
We fit the spectra in the 1--10\,keV energy range. We would
like to notice that similar parameter fits are found when fitting 
in the 0.3--10\,keV and 2--10\,keV energy ranges.

The column density ($N_{\rm H}$) of the gas along the line-of-sight
absorbing the X-ray photons was fitted using the {\tt XSPEC} model
{\tt wabs} \citep{MMc83}, which uses the (old) solar abundances of
\cite{Anders1982}. We notice that significant revisions of the solar
abundances have been made recently \citep[see e.g.,][for a
  review]{Asplund05}, leading to a decrease of the metal abundances,
in particular of carbon and oxygen, which are the main contributors to
the photoionization cross section above 0.3 and 0.6\,keV. Therefore,
using these revised solar abundances would increase the
absolute derived value of column density by about
50\% \citep{Grosso07}, with a negligible impact on the derived value of the photon index.
We also include the effect of the dust scattering, which deviates
soft X-ray photons from our line-of-sight, using P.\ Predehl's XSPEC {\tt scatter}
 model \citep{Predehl95}. 
The {\tt scatter} model multiplies the intrinsic spectrum by 
$\exp{(-\tau_{\rm sca})}$, where $\tau_{\rm sca}$ is the scattering optical depth, 
  which is given by $\tau_{\rm sca}=0.087\times (A_{\rm V}/{\rm
  mag})\times (E/{\rm 1\,keV})^{-2}$ \citep{Predehl95}. 
 From the $K$-band extinction value of
 2.8$\pm$0.2\,mag toward Sgr\,A* obtained using {\sl VLT/SINFONI}
 \citep{Eisenhauer05} and the 
 extinction law of \cite{Rieke85}, we inferred a dust visual
 extinction value of A$_{\rm V}$= 25.0$\pm$1.8\,mag.  

\begin{table}[!Ht]
\caption{Best fit parameters (using $W$ statistic, see text for details)
 of the EPIC flare spectra for
 absorbed pegged power-law (pow),  bremsstrahlung (brems), 
 and black-body (bb) models, taking into account dust scattering
 (assuming A$_{\rm v}$=25\,mag). The errors are given at the 90\%
 confidence level.
}
\label{tab:fitCstat}
\begin{tabular}{@{\ }clccccr@{\ }}
\hline
\hline
\noalign {\smallskip}
{\small Flare}   &  {\small Model}   & {\small $N_{\rm H}^{\rm \,(a)}$}
    & {\small $\Gamma$/kT$^{\rm (b)}$}  &{\small $F_{2-10\,{\rm keV}}^{\rm
   mean~(c)}$}   & {\small C/d.o.f.}\\
\noalign {\smallskip}
\noalign {\smallskip}
\hline
\noalign {\smallskip}
\multicolumn{6}{c}{April 2007}\\
\noalign {\smallskip}
\hline
\noalign {\smallskip}
$\#2$ &  pow          & 12.3$^{+2.1}_{-1.8}$     &
2.3$^{+0.3}_{-0.3}$  & 16.1$^{+3.1}_{-2.2}$ & 2560/2998 \\ 
\noalign {\smallskip}
        &brems       &  10.8$^{+1.6}_{-1.4}$     &
6.9$^{+3.1}_{-1.7}$  & 13.7$^{+4.3}_{-2.9}$ & 2559/2998 \\ 
\noalign {\smallskip}
        &bb       &   6.6$^{+1.3}_{-1.2}$     &
1.5$^{+0.1}_{-0.1}$  & 9.7$^{+0.7}_{-0.7}$ & 2562/2998 \\ 
\noalign {\smallskip}
\hline
\noalign {\smallskip}
$\#3$+$\#4$+$\#5$&  pow          &  8.8$^{+4.4}_{-3.2}$     &
1.7$^{+0.7}_{-0.6}$  & 5.0$^{+1.8}_{-1.0}$ & 2117/2998 \\ 
\noalign {\smallskip}
 &brems       &  8.3$^{+3.5}_{-2.4}$     & $\geq$ 7.0    &
 4.8$^{+2.6}_{-0.8}$ & 2117/2998 \\ 
\noalign {\smallskip}
 &bb       & 4.3$^{+2.8}_{-2.0}$     &
1.9$^{+0.4}_{-0.3}$  & 3.6$^{+0.7}_{-0.6}$ & 2117/2998 \\ 
\noalign {\smallskip}
\hline
\noalign {\smallskip}
\multicolumn{6}{c}{October 3, 2002}\\
\noalign {\smallskip}
\hline
\noalign {\smallskip}
 &  pow          & 12.3$^{+1.6}_{-1.5}$     &
2.2$^{+0.3}_{-0.3}$  & 25.3$^{+3.6}_{-2.7}$ & 2728/2998 \\ 
\noalign {\smallskip}
 &brems       &   10.7$^{+1.2}_{-1.1}$     &  8.2$^{+3.3}_{-1.9}$ & 21.9$^{+5.2}_{-3.6}$ & 2729/2998 \\ 
\noalign {\smallskip}
 &bb       & 6.4$^{+1.0}_{-0.9}$     & 1.6$^{+0.1}_{-0.1}$  & 15.6$^{+0.9}_{-0.8}$ & 2741/2998 \\ 
\noalign {\smallskip}
\hline
\end{tabular}
\begin{list}{}{}
\item[$^{\mathrm{(a)}}$] $N_{\rm H}$ is given in units of
  10$^{22}$\,cm$^{-2}$.
\item[$^{\mathrm{(b)}}$] The black-body temperature is given in keV. 
\item[$^{\mathrm{(c)}}$] Mean unabsorbed fluxes for the flare
 period in the 2--10\,keV energy range 
 in units of 10$^{-12}$\,erg\,cm$^{-2}$\,s$^{-1}$.
\end{list}
\end{table}

\begin{figure}[Ht!]
\includegraphics[angle=-90,width=\columnwidth]{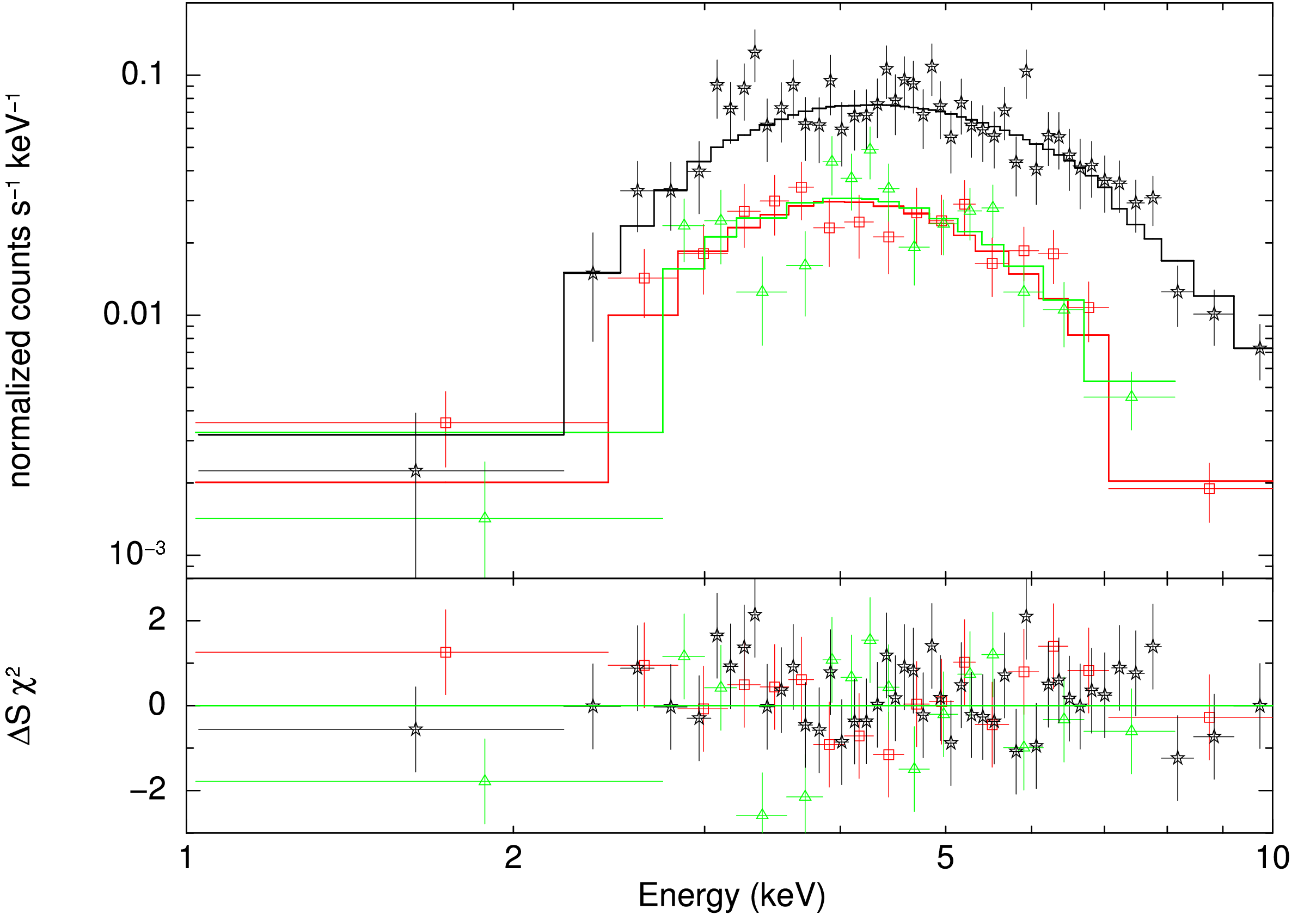}
\caption{{\sl XMM-Newton} EPIC spectra of the flare $\#2$ observed
  during the April 2007 observation campaign. 
 The spectrum for the non-flaring period with the same extraction region
  is used as background. 
Squares, triangles and stars indicate MOS1 (red), MOS2 (green), and
  pn (black) data, respectively. 
The lines show the best-fit model using an absorbed power-law model
  taking into account dust scattering (see Table~\ref{tab:fit} for
  the best fit parameter values using $\chi^{2}$ statistic; {\tt pow} model). The
  fit residuals are shown in the bottom panel.}
\label{fig:plfit}
\end{figure}

We fitted the spectra in turn with the following continuum models: 
 a pegged power-law model ({\tt pegpwrlw}\footnote{We used the
 {\tt pegpwrlw} (pegged power-law) model in which the unabsorbed flux 
over the energy range of the fit
 is used as normalization. This allows the photon index and the
 unabsorbed flux to be fit as independent parameters, and to 
 derive directly  the uncertainty on the unabsorbed flux.} model in
 XSPEC), a bremsstrahlung model, and a black-body model. 
 The parameters were tied between MOS1, MOS2, and pn spectra. 
Table~\ref{tab:fitCstat} gives  for these
models the best fit parameters and errors at the 90\%
 confidence level. The best-fit parameters of flare $\#2$ are rather well
constrained for the three models. 
For the absorbed power-law model we found  
$N_{\rm H}$=12.3$^{+2.1}_{-1.8}$$\times$
10$^{22}$\,cm$^{-2}$ and $\Gamma$=2.3$\pm$0.3. As reported
in Table~\ref{tab:fit}, very similar  best fit parameters are
obtained using the $\chi{^2}$ statistic ($N_{\rm H}$=12.8$^{+2.5}_{-2.1}$$\times$
10$^{22}$\,cm$^{-2}$ and $\Gamma$=2.3$\pm$0.4). 
For illustration purpose, we report in Fig.~\ref{fig:plfit} the
corresponding background-subtracted binned EPIC spectra
 and the data/model fit residuals obtained using the $\chi{^2}$ statistic. 
The presence of a narrow Fe\,K${\alpha}$ Gaussian emission lines 
($\sigma$=10\,eV) is not statistically required   
 ($\Delta \chi^{2}<$1 for one additional parameter). 
 We find at 6.4\,keV, 6.7\,keV,  and 7.0\,keV 
  upper limits on the equivalent widths of   
 154\,eV, 135\,eV, and 145\,eV, respectively. 
We also test for spectral variation index during the flare. First, splitting
 the spectra into two equal parts (as done in \citealt{P03c}), 
we obtained $N_{\rm H}$=13.0$^{+2.9}_{-2.5}$$\times$
10$^{22}$\,cm$^{-2}$, $\Gamma$=2.5$^{+0.5}_{-0.4}$, and $N_{\rm H}$=11.6$^{+3.0}_{-2.6}$$\times$
10$^{22}$\,cm$^{-2}$, $\Gamma$=2.1$\pm$0.5, for the rising and
 decreasing phases, respectively. These values are consistent within the
 error bars.  
If we tied the values of the column density (i.e., if we assume that it
 is constant during the flare), we found  $\Gamma$=2.4$^{+0.4}_{-0.3}$ and
 $\Gamma$=2.2$\pm$0.4, for the rising and  decreasing phases,
 respectively. Therefore, in both case, there is no significant (at a
 90$\%$ confidence level) spectral variations during the flare,
 consistent with the lack of spectral variability inferred above from the
 hardness ratio. The same conclusion is found when splitting the spectra
 into three parts: rising, top, and decreasing phases.
The flare $\#2$ data are also very well fitted by a bremsstrahlung and
 black-body continuum model with a temperature of about 6.8--6.9\,keV and
 1.5\,keV respectively (Tables~\ref{tab:fitCstat} and \ref{tab:fit}).  
 The column density value is, as expected, dependent of the
 continuum shape. 
The similar column density values found for power-law and
 bremsstrahlung continua, are significantly higher to the value of
 4.5$\times10^{22}$\,cm$^{-2}$ inferred by (dust) visual extinction
 ($A_{\rm V}\sim25$ mag), while found only slightly higher 
in case of a black body like shape. 

The best fit parameters for the sum of the weak flares
(\#3+\#4+\#5) are much less constrained than those of flare
\#2, but are consistent with the brightest flares within the error bars (i.e., at
90$\%$ of confidence level). 

Using a slightly higher dust visual extinction of $A_{\rm V}$=30\,mag
\citep[e.g.,][]{Rieke89}, as used for example in \cite{P03c},  
we found very similar parameter fit values. 
As an example, for an absorbed power-law model,
 we found for the flare $\#2$ $N_{\rm H}$=12.1$^{+2.1}_{-1.8}$$\times$
10$^{22}$\,cm$^{-2}$ and $\Gamma$=2.3$\pm$0.3 (using $W$ statistic).  

\begin{figure*}[!Ht]
\center
\begin{tabular}{@{}cc@{}}
\includegraphics[width=0.975\columnwidth]{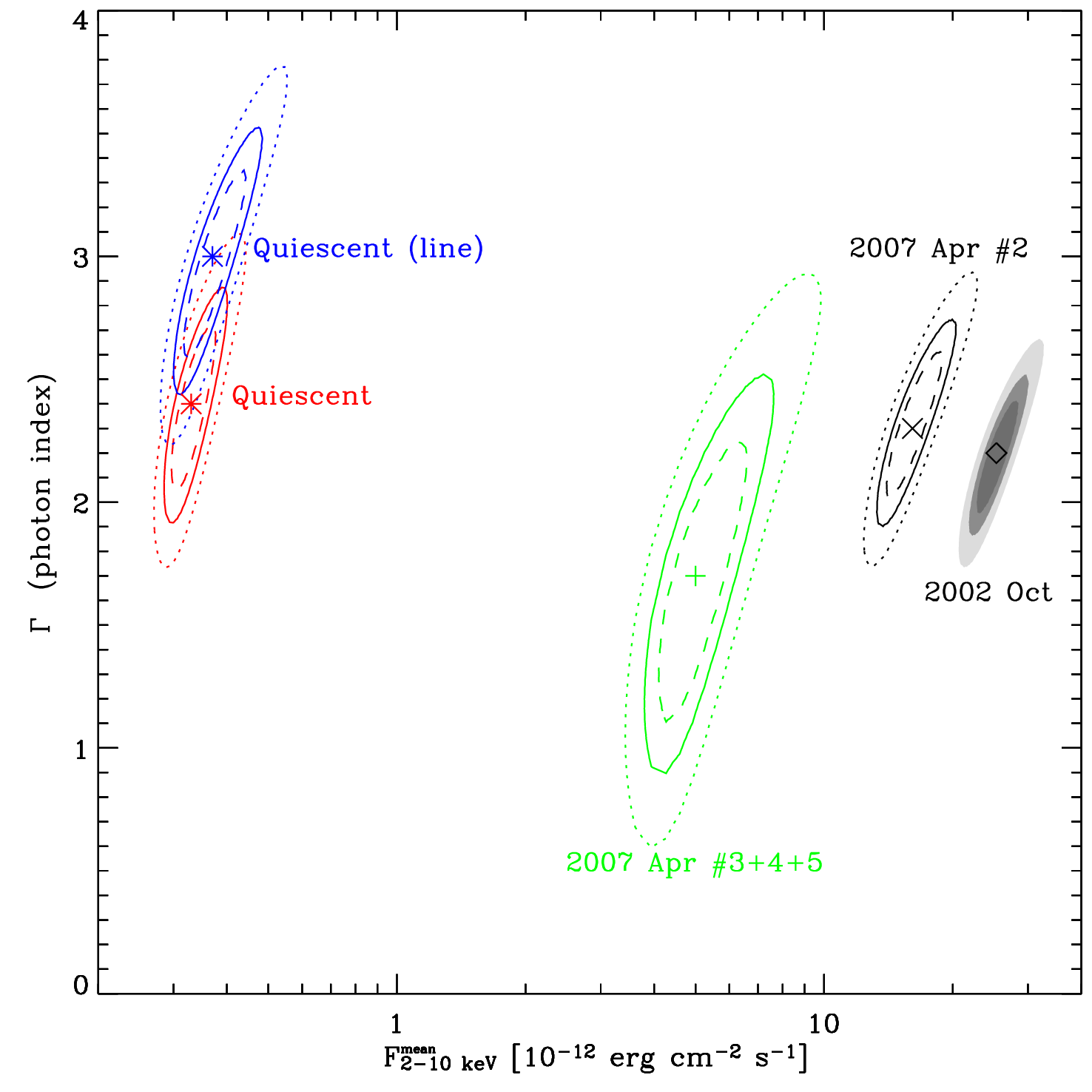} &
\includegraphics[width=0.975\columnwidth]{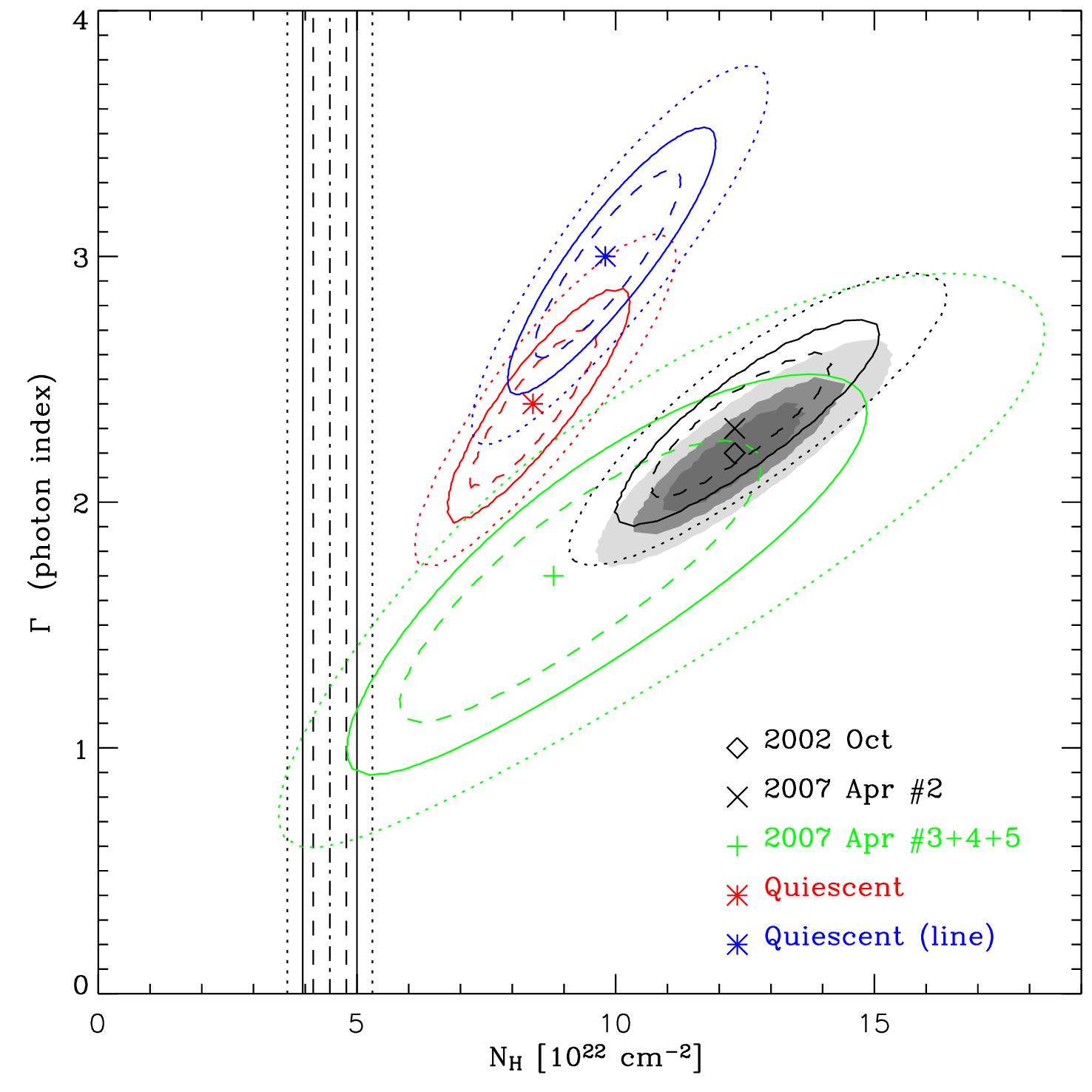}
\end{tabular}
\caption{Confidence regions  of the spectral parameters of the X-ray flares
  from Sgr\,A* for an absorbed power-law model taking into
  account the dust scattering. 
{\it Left panel}: photon index versus unabsorbed flux in the
  2--10\,keV energy range. The dashed, continuous, and dotted 
  contour levels correspond to confidence levels for two interesting
  parameters of 68$\%$, 90$\%$, 99$\%$, respectively (i.e., to $\Delta
  C$ = 2.3, 4.61, 9.21, respectively). The confidence regions 
  corresponding to the quiescent state of Sgr\,A*  
  are inferred from the spectral analysis of four archived Chandra
  observations, using the same spectral models (red
  contour levels) or including a Gaussian emission line (blue contour levels). See
  text for details. 
{\it Right panel}: photon index versus column density. The
  dashed-dotted vertical line indicates the column density
  (4.5$\times10^{22}$\,cm$^{-2}$) derived
  from the $K$-band extinction value toward Sgr\,A*  
  (2.8\,mag with one-sigma uncertainty of 0.2\,mag;
  \citealt{Eisenhauer05}) using the extinction law of 
  \cite{Rieke85} and the empirical relation of \cite{Predehl95}; 
   the dashed, continuous, and dotted vertical lines correspond
  to 68$\%$, 90$\%$, 99$\%$ confidence levels, respectively.} 
\label{fig:contourplots}
\end{figure*}

 The confidence regions of the photon index versus
the unabsorbed flux and the column density for the absorbed power
law model including dust scattering ($A_{\rm V}$=25\,mag) will be discussed in the
following section (see Fig.~\ref{fig:contourplots}).

\section{Comparison of the spectral parameters of the X-ray flares
  observed up-to-now with XMM-Newton}\label{sec:comparison}

Until now, the comparisons of X-ray flares from Sgr\,A* published in
  the literature have been based on spectral parameters obtained
  with different methods (e.g., different version of the SAS and CCF,
  different definition of flare time interval, scattering effect taken into
  account or not, etc). Here, we compare the spectral properties of the April
  2007 flares (\#2 and \#3+\#4+\#5) with those of the rapid X-ray flares
  observed previously with {\sl XMM-Newton} that we obtained using
  exactly the same method.

\subsection{Reprocessing and spectral analysis of X-ray flares
  previously observed with XMM-Newton}

We have reprocessed the observations of October 3,
2002 (orbit 516, ObsID: 0111350301; \citealt{P03c}) and March 31, 2004 (orbit 789, ObsID:
0202670601; \citealt{Belanger05}) with the SAS version 7.1 and
 the latest CCF (see Appendix~\ref{app:appB}). 
We did not reprocess the X-ray flare observed on September 4, 2001,
 because only a part of the beginning of the flare was observed
\citep{Goldwurm03}. 
We did not consider here the X-ray flare observed on August 31, 2004
\citep{Belanger05} because it was contaminated by an X-ray transient
  in outburst, located at only 2.9\arcsec-South from 
Sgr\,A*, displaying dips/eclipses at this epoch \citep{Muno05,P05b},
 which precludes the secure determination of
the spectral parameters in this case. 

The reprocessed light curves of the October 2002 and March 2004 flares 
are shown in the online Fig.~\ref{fig:zoom_anteriorflare}. 
The time intervals of the flares
have been defined using the same criteria used in $\S$\ref{sec:lc},
 and reported in the online Table~\ref{tab:flaretime_2}. 
As done above, the 
  10$^{\prime\prime}$-radius extraction region has been optimized in
each instrument by determining the position of Sgr\,A* during the time interval of the flare.
The flare time interval found here for the October 2002 flare is very
similar to the one used in \cite{P03c}. While, the flare time interval for
the March 2004 found here using a time bin of 150\,s 
is shorter (~1200\,s) than the time interval determined in
\cite{Belanger05} using a much large time bin ($\sim$~500\,s), 
  which contained a significant contribution of the non-flaring
  state. 

We fitted the corresponding EPIC background-subtracted unbinned
spectra with an absorbed power
law model including dust scattering ($A_{\rm V}$=25\,mag) 
 and the $W$ statistic as for the
 flares $\#$2 and $\#$3+$\#$4+$\#$5 ($\S$\ref{sec:spectral}). 
In the October 2002 observation, the non-flaring period before
 the flare is too short ($\sim$10\,ks) to allow a correct 
 determination of the complex shape of the background spectrum of the non-flaring level, 
which is crucial when using $W$ statistic to fit the data. Therefore, we have used the
 longer observation of February 2002 (orbit 406, ObsID: 0111350101)
 where no Sgr\,A* flare were observed. Indeed, the non-flaring spectrum observed on
 February 2002 has the same shape and flux than the non-flaring
 spectrum observed on October 2002 before the flare, but it much more
 well constrained. We notice that using the $\chi^{2}$
 statistic, using background spectra based either on the October 2002
 observation or on the February 2002 observation, no impact on the
 parameter fits are found.

For the October 2002 flare the best fit parameters are reported in Table~\ref{tab:fitCstat}. 
The online Fig.~\ref{fig:plfit_oct02} shows for illustration purpose   
the corresponding background-subtracted binned spectra
and best-fit model for an absorbed power-law model using the $\chi^{2}$ statistic.  
As for the flare $\#2$, fits with the $\chi^{2}$ statistic give very similar
 parameter fits (see Table~\ref{tab:fit}). 
Although, the central value of the photon index, $\Gamma$=2.2 ($\pm$0.3), is slightly
lower than the value, $\Gamma$=2.5 ($\pm$0.3), previously reported by
\cite{P03c}, both values are consistent when taking into account 
the error bars at the 90$\%$ confidence level. 
For this flare, both the position of Sgr\,A* during the flare and the flare time interval 
 used in \cite{P03c} were very similar to those used here, and
 then, are not responsible for the small difference in the central value of the
 photon index found here. 
Since the first analysis of the brightest flare
by \cite{P03c} with the SAS version 5.4, the cross-calibration
between MOS and pn has been substantially improved (see e.g.,
\href{http://xmm.vilspa.esa.es/docs/documents/CAL-TN-0018.ps}{XMM-CAL-TN-0018}).
We checked that using the same extractions region, flare and
 background time intervals as used in \cite{P03c} and the new calibration used in this work, 
we retrieved $\Gamma\sim$2.2. 

For the moderate flare observed on March 31, 2004, we found 
$N_{\rm H}$=4.5$^{+3.4}_{-2.2}$$\times$10$^{22}$\,cm$^{-2}$ and
$\Gamma$=0.8$\pm$0.6 (C/d.o.f.=1860/2998). Similar values are obtained
  with the $\chi^{2}$ statistic ($N_{\rm H}$=4.7$^{+4.6}_{-3.0}$$\times$10$^{22}$\,cm$^{-2}$ and
$\Gamma$=0.9$^{+1.0}_{-0.9}$, $\chi^{2}$/d.o.f.=18.4/21; see Fig.~\ref{fig:plfit_march04}). 
This central photon index value is smaller than the one reported by
\cite{Belanger05} ($\Gamma$=1.5$^{+0.5}_{-0.6}$ with the error bars
calculated at the 68$\%$ confidence level), though compatible within
the large error bars.
The difference is mainly due to the method used here: 
a better determination of the flare time interval, which
  helps to decrease the contribution of the non-flaring level; a
spectral fit done without binning the spectra when using the $W$ statistic, 
in order to not loose spectral information and then to not bias the
parameter fit result; the dust scattering effect taken into account. 
A closer look to the fit of the March 2004 spectra show that when the
instruments are fitted separately, very different central fit values are
obtained (although compatible within the very large error
bars). Moreover, the inferred column density value is lower than the
one determined during the quiescent state (see below) and therefore appears
unphysically low. 
When the column density is fixed to the value
found for the brightest flares, the spectral index is
1.8$\pm$0.4 using the $W$ statistic (C/d.o.f=1872/2999) and
2.3$\pm$0.6 using the $\chi^{2}$ statistic ($\chi^{2}$/d.o.f.=24/22), 
which is consistent within the error bars with the spectral
  index value of the brightest flares. Therefore, we conclude
  that the spectral properties of this moderate flare are poorly
  constrained, and we will not considered it in our comparison
of the flare properties.

\subsection{Confidence regions of the spectral parameters}

We show in the left and right panels of Fig.~\ref{fig:contourplots}
the contour plot of the confidence regions of the photon index versus
the unabsorbed flux and the column density, respectively, at the confidence levels of
68\%, 90\%, and 99\% (corresponding to $\Delta C$=2.3,
4.61, and 9.21, respectively, for two interesting parameters). 
The fluxes of the flares span a large range, i.e., about a factor
5. The flare $\#$2 has a flux lower than the October 2002 
flare at the 90\% confidence level, but has similar best fit values 
 of column density and spectral index.    
Due to a lower S/N, the physical parameters of the sum of the
moderate flares (\#3+\#4+\#5) are less constrained, but has column
density and spectral index values compatible (within the error bars) 
with those of the bright flares. 

The column density values during the flares are at least two 
  times higher than the value expected from the 25\,mag of (dust)
  visual extinction toward Sgr\,A*, i.e.,
  4.5$\times10^{22}$\,cm$^{-2}$. 
To know whether this excess of (gas) column density is
related to the flare phenomena itself, we need to estimate
  the column density of the quiescent spectrum of Sgr\,A* with the
  same spectral model. However, the quiescent emission of Sgr\,A* can only be
  resolved with {\sl Chandra}. The spectral properties of the
  quiescent emission of Sgr\,A* was reported by \cite{Baganoff03},
  based on the first {\sl Chandra} observations of the Galactic
  Center obtained in 1999 with a total effective exposure of about
  41\,ks. More {\sl Chandra} observations of Sgr\,A* are now available
  in the archives. Starting from the level 2 of the archive event lists, we used {\tt
  CIAO 4.0} and {\tt CALDB 3.4.2} to produce {\sl Chandra} X-ray light curves
  of Sgr\,A*\footnote{See
  \href{http://asc.harvard.edu/ciao/threads/index.html}{Science threads
  of {\tt CIAO 4.0}}.}. We selected four {\sl Chandra} observations
  where Sgr\,A* displayed no flares\footnote{Namely ObsID 3665, 4683,
  5950, and 5951 with exposure times of about 90, 50, 48, and 44\,ks,
  respectively.}. We extracted the corresponding quiescent spectra
  following the method of \cite{Baganoff03}; we refined the background
  10$\arcsec$-radius extraction region by excluding the candidate
  pulsar wind nebula G359.95-0.04.

We have performed the simultaneous spectral analysis of these four
quiescent spectra of Sgr\,A* obtained with {\sl Chandra}.    
First, we fit the data with an absorbed power-law model using the
$\chi^{2}$ statistic and a spectral binning of 10 counts per bin as
done in \cite{Baganoff03}. When the dust
scattering is not taken into account, we found $N_{\rm H}$=9.7$^{+1.8}_{-1.5}\times$
10$^{22}$\,cm$^{-2}$ and $\Gamma$=2.5$^{+0.5}_{-0.4}$
($\chi^{2}$/d.o.f.=143/130) and an unabsorbed 2--10\,keV flux of
2.8$^{+0.6}_{-0.4}\times$10$^{-13}$\,erg\,cm$^{-2}$\,s$^{-1}$. These values 
are similar to those reported by \cite{Baganoff03} but are better
  constrained. An excess of emission is seen near 6.5\,keV,
  therefore we add a Gaussian emission line and found $N_{\rm
    H}$=11.1$^{+2.1}_{-1.9}\times$10$^{22}$\,cm$^{-2}$ and
  $\Gamma$=3.0$^{+0.5}_{-0.6}$ and an unabsorbed 2--10\,keV flux of 
3.2$^{+0.6}_{-0.7}\times$10$^{-13}$\,erg\,cm$^{-2}$\,s$^{-1}$, E$_{\rm
  line}$=6.61$^{+0.10}_{-0.10}$\,keV, $\sigma_{\rm
  line}$=0.15$^{+0.10}_{-0.12}$\,keV, and $EW_{\rm
  line}$=1.1$^{+0.4}_{-0.5}$\,keV ($\chi^{2}$/d.o.f.=123/127). 
As found by \cite{Baganoff03}, we found a steeper power-law
slope when an emission line is added to the fit. In this work, thanks
to a much longer exposure time, we are able to constrain the line
energy, which is consistent with a He-like iron line. 
\begin{figure}[!t]
\centering
\includegraphics[angle=-90,width=\columnwidth]{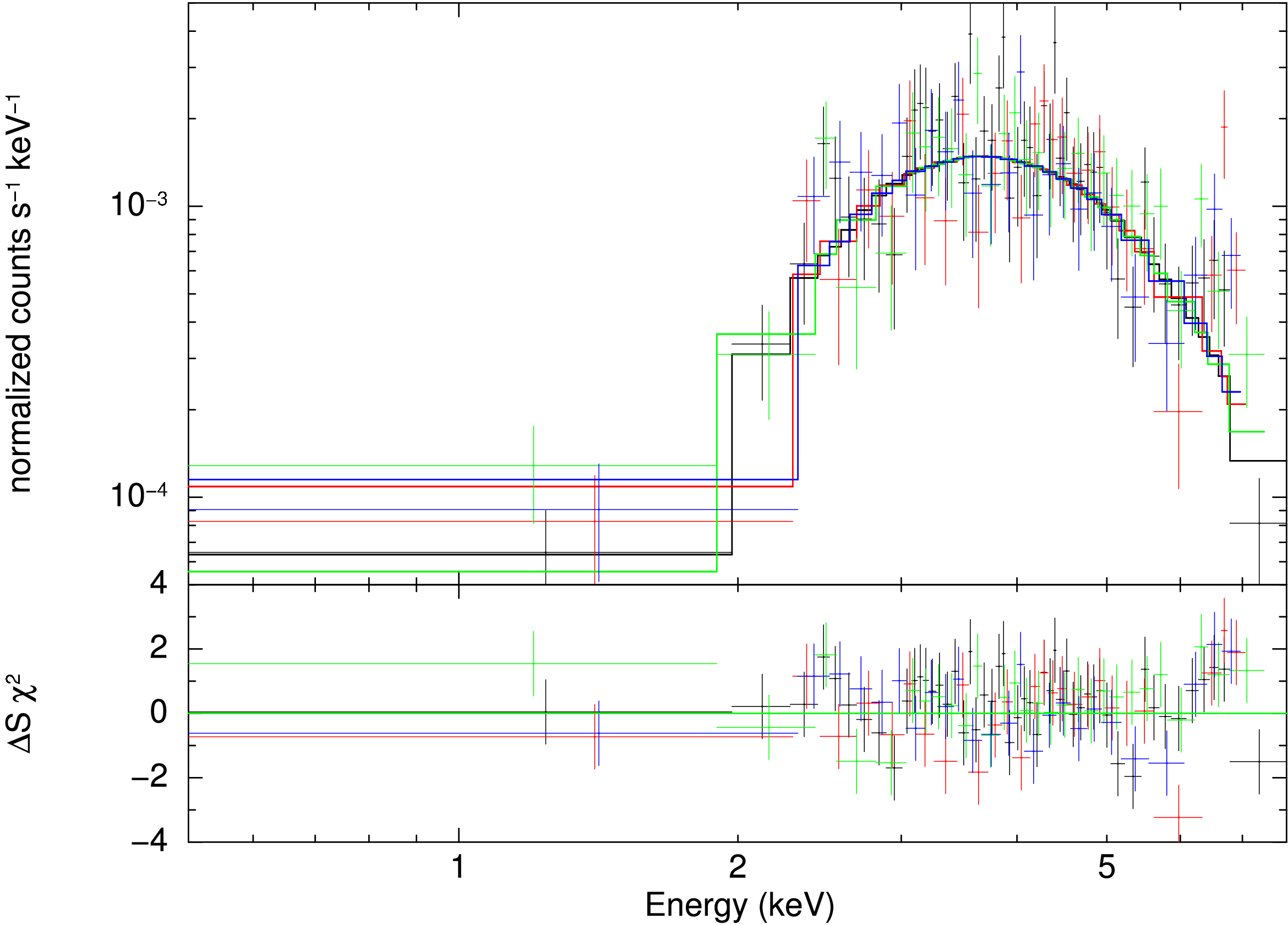}
\caption{{\sl Chandra} quiescent spectra of Sgr\,A*. Black, red, green,
  and blue data points correspond to ObsID 3665,  4683, 5950, and
  5951, respectively. The lines show the best-fit model ($\chi^{2}$
  statistic) using an absorbed power-law model taking into account the
  dust scattering. The fit residuals are shown in the bottom panel.
}
\label{fig:plfitcxo}
\end{figure}
For illustration purpose, we show in Fig.~\ref{fig:plfitcxo}
  the background-subtracted binned {\sl Chandra} spectra and the
  data/model fit residuals obtained for an absorbed power-law model
  taking into account the dust scattering effect with the $\chi{^2}$
  statistic. 
To perform a homogeneous comparison with the spectral analysis
of the {\sl XMM-Newton} X-ray flares reported in this work, 
we then fit the unbinned {\sl Chandra} quiescent spectra using the $W$ statistic
and we took into account dust scattering effect. We found $N_{\rm
  H}$=8.4$^{+1.4}_{-1.3}$$\times$ 10$^{22}$\,cm$^{-2}$,
  $\Gamma$=2.4$^{+0.4}_{-0.4}$ ($C$/d.o.f.=1825/2457), corresponding
to an unabsorbed 2--10\,keV flux of
  3.3$^{+0.5}_{-0.4}\times$10$^{-13}$\,erg\,cm$^{-2}$\,s$^{-1}$
 and a luminosity of 2.4$^{+0.4}_{-0.3}\times$10$^{33}$\,erg\,s$^{-1}$ 
(assuming d=8\,kpc), similar to the value reported by \cite{Baganoff03}.  
The contour plots corresponding to this fit
  are reported on Fig.\,\ref{fig:contourplots}. 
 If we add a Gaussian emission line we found $N_{\rm H}$=9.8$^{+1.8}_{-1.5}\times$ 
10$^{22}$\,cm$^{-2}$, $\Gamma$=3.0$\pm$0.5, E$_{\rm
  line}$=6.67$^{+0.05}_{-0.07}$\,keV, 
$\sigma_{\rm  line}\leq$0.19\,keV, and $EW_{\rm
  line}$=0.9$^{+0.4}_{-0.3}$\,keV (C/d.o.f.=1781/2454). 
 As shown in Fig.~\ref{fig:contourplots}, the column density and power-law
  index values found for the quiescent state (Gaussian line added or
  not to the fit), are compatible, within the error bars calculated at
  90\% confidence level for two interesting parameters, 
with those found for the bright flares.  

\section{Summary and discussions}\label{sec:discussion}

Here, we reported the data analysis of the {\sl XMM-Newton} April 2007
campaign observation of Sgr\,A* (three observations with a total of
exposure of $\sim$230\,ks). 
We observed four X-ray flares occurring within only half a day on
April 4: one bright flare ($\#2$, with a detection level of $\sim$21$\sigma$) 
followed by three moderate ones
($\#3$, $\#4$, and $\#5$, with a detection level of $\sim$6$\sigma$,
$\sim$6$\sigma$, and $\sim$8$\sigma$, respectively). 
 As well a weak flare ($\#1$, with a detection level of $\sim$3$\sigma$)
 was observed on April 2, 2007. 
The flare $\#2$ is the second brightest X-ray flare detected so far
from Sgr\,A* and has a similar duration ($\sim$3\,ks) to that of the
October 2002 brightest flare 
\citep{P03c}. Its light curve is almost symmetrical but without no significant
 short-time scale drop (i.e., about 50\% of flux decrease) 
contrary to that reported for the October
 2000 flare \citep{Baganoff01} and the October 2002 flare
 \citep{P03c}. However, for the flare $\#$2 we
cannot rule out a moderate drop in the X-ray light curve.
A similar group of three moderate flares were already observed with
  {\sl XMM-Newton} on 2004 March 31 \citep{Belanger05}, but no such
  preceding bright flare was observed. 

This is the first time that a
  such level of X-ray flaring activity from Sgr\,A*, both in amplitude and
  frequency, is reported. 
 Observations such as those reported in this paper can eventually
constrain flare models in Sgr A*, perhaps even ruling some out.
For example, the quick succession of several events separated by
only a few hours, might argue against a disruption mechanism
that relies on the temporary storage of mass and energy, if all
  the corresponding accretion energy is released during the flare. 
The accretion rate $\dot M$ in this system (some estimates place it as low
as $10^{16}-10^{17}$ g s$^{-1}$; see \citealt{Melia07}) might not produce
a transient accumulation of mass $\Delta M$ between flares of
sufficient magnitude for $\eta GM\Delta M/ 3R_S$ to provide the
observed outburst power. In this expression, $R_S$ is the Schwarzschild
radius, and $\eta$ is the (poorly known) radiative efficiency
in Sgr A*, believed to be at most a few percent. If instead the
flares are due to a magneto-rotational instability, then the energy
liberated during the flare must also be accumulated over the short
inter-burst period. The low $\dot M$, from which the energy is
derived, might argue against this type of mechanism as well.

 On the other hand, if the flare arises from the infall of a clump
of gas \citep[e.g.,][]{Liu06,Tagger06} 
then there would be less restriction on how often these could come
in. \cite{Genzel03} have shown that the total energy release $\geq$
10$^{39.5}$ erg during a flare requires a gas accreted mass of a few times
10$^{19}$\,g (assuming a radiation efficiency of $\sim$10$\%$), i.e.\
comparable to that of a comet or a small asteroid. Recently, 
\cite{Cadez06} have argued that the flares could be
produced by tidal captures and disruptions of such small bodies. 
 The comet/asteroid/planetesimal idea for depositing the additional mass and energy
to initiate a flare is attractive. 
The distance from Sgr A* at which such a small body 
would get tidally disrupted is the Roche radius, which is for a rigid
body:
\[
 \frac{R_{\cal R}}{R_{\rm S}} = 13.2 \times \left(\frac{M_{\rm BH}}{4 \times 10^6
 M_\odot}\right)^{-2/3} \times \left(\frac{\rho_p}{{\rm
 1\,g\,cm^{-3}}}\right)^{-1/3}\, , 
\]
 where $M_{\rm BH}$ is the black hole mass and 
$\rho_p$ is the density of the rigid body. Thus, for a black-hole
mass of $4\times 10^6\;M_\odot$ and a density of 1\,g\,cm$^{-3}$, this
corresponds to $13.2\,R_{\rm S}$, 
in good agreement with the size of the region where the flares are
thought to occur. 
The flaring rates would then depend on processes occurring much farther out, so the
fact that so many flares are seen so close together on some days,
and much less frequently at other times, would simply be due to
stochastic events. However, it would still be difficult to
distinguish between a compact emission region and emission within
a jet (see, e.g., \citealt{Markoff01}), 
since these disruption events could still end up producing an
ejection of plasma associated with the flare itself.\\

  Four of the five X-ray flares observed during this campaign have
 a simultaneous NIR flare counterpart: flares $\#1$, $\#4$, and $\#5$
 were observed with HST/NICMOS (Yusef-Zadeh et al.\ 2008, in prep.);
 and flare $\#2$ was observed with VLT/NACO (Dodds-Eden et al.\ 2008,
 in prep.). The flare $\#3$ has not been covered by a simultaneous
 NIR observation.  
This strengthens the relationship observed up to now between X-ray and
 NIR flares when there is a simultaneous X-ray/NIR observation coverage: 
all X-ray flares have an NIR flare counterpart, while all NIR 
flares are not each time associated with an X-ray flare counterpart
 \citep[e.g.,][]{Eckart04,Eckart06,Yusef06,Hornstein07,Marrone08}.

 We have made the first detailed comparison of X-ray flare properties
 observed with {\sl XMM-Newton} (this April 2007 campaign and previous
 observations) based on a fully self-consistent analysis approach,
 namely we use the same SAS version with the latest
 calibration, the same definition of the flare interval 
(preventing from significant contribution of the
 non-flaring level), an optimized extraction region for each instrument, 
 and we took into account the dust scattering effect.  
We found that the physical parameters of the flare $\#2$, $N_{\rm H}$=12.3$^{+2.1}_{-1.8}$$\times$
10$^{22}$\,cm$^{-2}$ and $\Gamma$=2.3$\pm$0.3, are well constrained
and are very similar to that of the October 2002 (brightest) flare, $N_{\rm
  H}$=12.3$^{+1.6}_{-1.5}$$\times$10$^{22}$\,cm$^{-2}$ and $\Gamma$=2.2$\pm$0.3. 
The spectral parameter fit values of the sum of the three moderate 
 flares following the bright flare observed on April 4, 2007, 
$N_{\rm H}$=8.8$^{+4.1}_{-3.2}$$\times$ 10$^{22}$\,cm$^{-2}$ and $\Gamma$=1.7$^{+0.7}_{-0.6}$,  
while lower, are compatible within the error bars with those of the bright flares. 
The column density found during the bright flares is at least two times
higher than the value expected from the (dust) visual extinction
 toward Sgr\,A* ($A_{\rm V}\sim25$ mag; \citealt{Eisenhauer05}), i.e., 
 4.5$\times10^{22}$\,cm$^{-2}$. However, our fitting of the Sgr\,A* quiescent
 spectra obtained using four {\sl Chandra} observations for a total
 observation time of $\sim$ 230\,ks, taking into account the dust
 scattering, shows that a column density excess is already
 present during the non-flaring phase.  
 One possible explanation, as already pinpoint by \cite{Maeda02}, 
 is that such column density excess in the
line-of-sight of Sgr\,A* could be due to the warm (dust free) gas
 associated with the ionized gas halo, \object{Sgr\,A West} extended,
 suggested by the turnover 
 absorption observed at 90\,cm, embedded the Sgr\,A complex
  \citep[e.g.,][]{Pedlar89,Anantharamaiah99,Yusef00}. 
\cite{Anantharamaiah99} inferred that this ionized halo has a dimension of about
4$\arcmin$ ($\sim$9pc) and an electron density of about 100--1000\,cm$^{-3}$. 
For a column excess of about 2.9$\times$10$^{22}$\,cm$^{-2}$ as found for
the Sgr\,A* quiescent state, this would correspond to a consistent electron
density of about 1300\,cm$^{-3}$. 
 However, in case the genuine continuum shape is curved compared
 to a power law and bremsstrahlung models, as for example a black body
 like shape, no significant column excess would be required. 

 On the basis of our current analysis we
conclude that the dichotomy of the spectral index between
moderate and bright flares (i.e., a correlation between photon
index and flux) suggested by \cite{Belanger05} 
 is not confirmed. However, to test this more thoroughly
we need further observations so as to extend the sampling of the flaring
activity seen from Sgr A* (and in particular those instances where
good signal to noise can be achieved through the co-adding of
weak to moderate flaring events).

In conclusion, this study establishes that the two brightest X-ray
flares observed so 
far from Sgr\,A* exhibited similar (well constrained) soft spectra. 
Therefore, any model proposed to explain the flaring behavior of Sgr\,A* must
take into account this observational X-ray spectral property.

\begin{acknowledgements}
The XMM-Newton project is an ESA Science Mission with instruments
and contributions directly funded by ESA Member States and the
USA (NASA). 
\end{acknowledgements}

\bibliographystyle{aa}
\bibliography{biblio}


\Online

\phantomsection 
\addcontentsline{toc}{chapter}{Online material}

\phantomsection 
\addcontentsline{toc}{chapter}{\appendixname}

\begin{appendix}
\section{Absolute astrometry of the bright X-ray flare}\label{app:astrometry}

To obtain an absolute astrometry of the bright X-ray flare, we
registered the positions of the soft X-ray sources in the third
observations on the {\sl Hipparcos} celestial coordinate system using
the X-ray counterparts of the Tycho-2 catalog's sources
\citep{hog00}. We matched the positions of the pn sources detected in
the 0.5--1.5\,keV energy band obtained with the SAS task {\tt edetect\_chain}
with the positions of Tycho-2 catalog's
sources for the epoch of our observation. We found 4 X-ray
counterparts of Tycho-2 sources using a cross-correlation radius of
$5\arcsec$. The X-ray counterpart of Tyc\,6840~020~1, lying on a
pn CCD gap, is not considered in the following. The offsets between
pn and Tycho-2 positions of these three reference sources shows a
systematic translation. The weighted mean offset is $-0\farcs7$ and
$0\farcs6$ in right ascension and declination,
respectively. Therefore, we correct the X-ray position by subtracting
this weighted mean offset. The residual RMS scatter in the corrected
X-ray positions of the reference sources, $0\farcs9$, gives the
astrometric uncertainty of the registration. This value is combined
with the positional error of the PSF fitting to derived the final
positional uncertainty. 
We list in Table~\ref{tab:tycho} the corrected X-ray positions of the Tycho-2
reference stars, the final positional uncertainties, and the distance
to the Tycho-2 optical positions. The absolute position of the X-ray
bright flare is $\alpha_{\rm J2000}= 17^{\rm h}45^{\rm m}40.0^{\rm
s}$, $\delta_{\rm J2000}=-29^{\circ}00^{\prime}28.6^{\arcsec}$ with a
one-sigma positional uncertainty of $1\farcs0$.

\begin{table}[!h]
\caption{X-ray counterparts of Tycho-2 reference stars.}
\label{tab:tycho}
\begin{center}
\begin{tabular}{@{}ccccc@{}}
\hline
\hline
\noalign {\smallskip}                       
Name & J2000 coordinates$^{\mathrm{(a)}}$ & Err.$^{\mathrm{(b)}}$ &  Name    & Dist.\\
XMMU J1745 &  $17^{\rm h}45^{\rm m}$                 & $\arcsec$ & Tyc~6840  & $\arcsec$ \\
\hline
25.7-285627 & $25.7^{\rm s}$ $-28^{\circ}56^{\prime}27.8^{\arcsec}$ & 1.3 & 666~1 & 0.9\\
43.8-291317 & $43.8^{\rm s}$ $-29^{\circ}13^{\prime}17.7^{\arcsec}$ & 1.7 & 334~1 & 0.9\\
43.9-290456 & $43.9^{\rm s}$ $-29^{\circ}04^{\prime}56.5^{\arcsec}$ & 1.1 & 590~1 & 0.2\\
\hline
\end{tabular}
\begin{list}{}{}
\item[$^{\mathrm{(a)}}$] Absolute position corrected from systematic translation.\\
\item[$^{\mathrm{(b)}}$] Final positional uncertainty obtained by
  combining the positional error of the PSF fitting with the
  astrometric uncertainty of the registration on Tycho-2 sources.\\
\end{list}
\end{center}
\end{table}
\end{appendix}

\begin{appendix}
\section[Chi-square statistic versus $W$ statistic]{$\chi{^2}$ statistic versus $W$ statistic}\label{app:stat}

 The flare spectra are rebinned from 1\,keV 
using {\tt grppha} to have at least 25 counts per spectral bin. 
We ignore data above 10\,keV, as for the fits with the W statistic
(see $\S$\ref{sec:spectral}). 
{\tt XSPEC} (version 12.4.0; \citealt{Arnaud96})
 was used to fit the background-subtracted spectrum. 
For the October 2002, as for the $W$ statistic spectral fit, we used as
background the longer February 2002 observation (orbit 406,
ObsID: 0111350101) where no Sgr\,A* flare were observed. 
If the non-flaring spectrum observed on October 2002 before the
flare is used as background, very similar parameter fits are obtained. 
 The errors and upper limits quoted correspond to 90$\%$ confidence
level for one interesting parameter.
The parameter fits (see Table~\ref{tab:fit}) are similar to that found using the $W$ statistic
for bright flares. For lower S/N flares the central values are different
but consistent within the error bars with that the values found using
$W$ statistic. The error bars are generally larger using the
$\chi{^2}$ statistic. 

\begin{table}[!Ht]
\caption{Best fit parameters (using $\chi^{2}$ statistic) of the EPIC flare spectra for
 absorbed pegged power-law (pow),  bremsstrahlung (brems), 
 and black-body (bb) models, taking into account dust scattering
 (assuming A$_{\rm v}$=25\,mag). The errors are given at the 90\%
 confidence level.
}
\label{tab:fit}
\begin{tabular}{@{\ }cclc@{\ }c@{\ }c@{\ }c@{\ }r@{\ }}
\hline
\hline
\noalign {\smallskip}
{\small Flare} & {\small Model}  & \multicolumn{1}{c}{\small $N_{\rm H}^{\rm \,(a)}$}
    & \multicolumn{1}{c}{\small $\Gamma$/kT$^{\rm (b)}$}  &\multicolumn{1}{c}{\small $F^{\rm mean (c)}_{2-10\,{\rm keV}}$}   & {\small $\chi^{2}$/d.o.f.} & \multicolumn{1}{c}{\small  $\cal{Q}^{\rm (d)}$} \\
\noalign {\smallskip}
\hline
\noalign {\smallskip}
\multicolumn{7}{c}{April 4, 2007}\\
\noalign {\smallskip}
\hline
\noalign {\smallskip}
$\#2$ & pow & 12.8$^{+2.5}_{-2.1}$     &
2.3$^{+0.4}_{-0.4}$  & 15.7$^{+3.6}_{-2.4}$ & 70.2/75 & 63.6$\%$  \\
\noalign {\smallskip}
        &brems       &  11.3$^{+1.9}_{-1.6}$     & 6.8$^{+3.7}_{-1.9}$  & 13.4$^{+5.2}_{-3.1}$ & 69.8/75 & 64.9$\%$  \\
\noalign {\smallskip}
        &bb       &   7.3$^{+1.6}_{-1.3}$     & 1.5$^{+0.1}_{-0.1}$  & 9.5$^{+0.8}_{-0.7}$ & 71.5/75 & 59.2$\%$  \\
\noalign {\smallskip}
\hline
\noalign {\smallskip}
 {\small $\#3$+$\#4$+$\#5$}$^{\rm (e)}$ &  pow & 6.3$^{+4.6}_{-2.8}$     &
1.2$^{+0.7}_{-0.6}$  & 4.1$^{+1.3}_{-0.7}$ & 29.9/32 & 57.3$\%$  \\
\noalign {\smallskip} 
 &brems       &   6.4$^{+2.3}_{-1.6}$     & $\geq$ 11  & 4.1$^{+6.6}_{-1.4}$ & 29.9/32 & 57.4$\%$  \\
\noalign {\smallskip}
 &bb       & 2.9$^{+2.6}_{-1.7}$     & 2.1$^{+0.6}_{-0.4}$  & 3.3$^{+1.2}_{-0.7}$ & 29.7/32 & 58.2$\%$  \\
\noalign {\smallskip}
\hline
\noalign {\smallskip}
\noalign {\smallskip}
\multicolumn{7}{c}{October 3, 2002}\\
\noalign {\smallskip}
\hline
\noalign {\smallskip}
 & pow         & 12.1$^{+1.7}_{-1.5}$     &
2.2$^{+0.3}_{-0.3}$  & 24.1$^{+3.6}_{-2.7}$ & 90.4/102 & 78.8$\%$  \\
\noalign {\smallskip} 
 &brems       &   10.6$^{+1.3}_{-1.1}$     &  7.9$^{+3.4}_{-1.9}$ & 20.9$^{+5.2}_{-3.6}$ & 90.9/102 & 77.6$\%$  \\
\noalign {\smallskip}
 &bb       & 6.5$^{+1.1}_{-1.0}$     & 1.6$^{+0.1}_{-0.1}$  & 14.9$^{+0.8}_{-0.8}$ & 98.1/102 & 59.0$\%$  \\
\noalign {\smallskip}
\hline
\end{tabular}
\begin{list}{}{}
\item[$^{\mathrm{(a)}}$] $N_{\rm H}$ is units of
  10$^{22}$\,cm$^{-2}$.
\item[$^{\mathrm{(b)}}$] The black-body temperature is given in keV. 
\item[$^{\mathrm{(c)}}$] Mean unabsorbed fluxes for the flare
 period in the 2--10\,keV energy range 
 in units of 10$^{-12}$\,erg\,cm$^{-2}$\,s$^{-1}$.
\item[$^{\mathrm{(d)}}$] $\cal{Q}$ is the null hypothesis probability,
  i.e., the probability of getting a value of $\chi^{2}$ as large or
  larger than observed if the model is correct.
\item[$^{\mathrm{(e)}}$] These parameter fits for flare
  $\#3$+$\#4$+$\#5$ are only given to
  illustrate the impact on both the central parameter
  fit values and the constraints on error bars, when using $\chi^{2}$
  statistics in case of low statistics, instead of $W$ statistics (see
  Table~\ref{tab:fitCstat} for comparison). 
\end{list}
\end{table}
\end{appendix}

\begin{appendix}
\section{Reprocessed light curves and spectra of the X-ray flares previously observed with XMM-Newton}\label{app:appB}

\begin{figure}[!ht]
\centering
\begin{tabular}{@{}cc@{}}
\includegraphics[width=0.46\columnwidth]{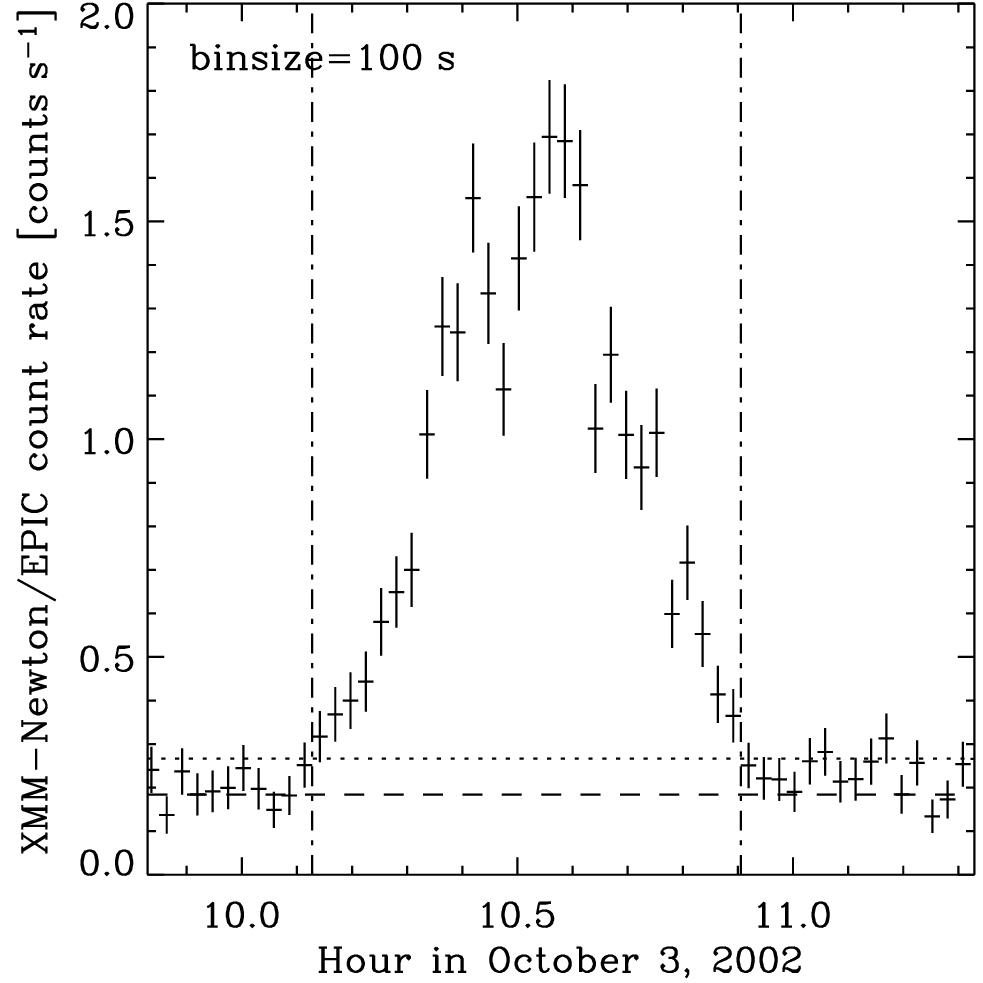} 
& \includegraphics[width=0.46\columnwidth]{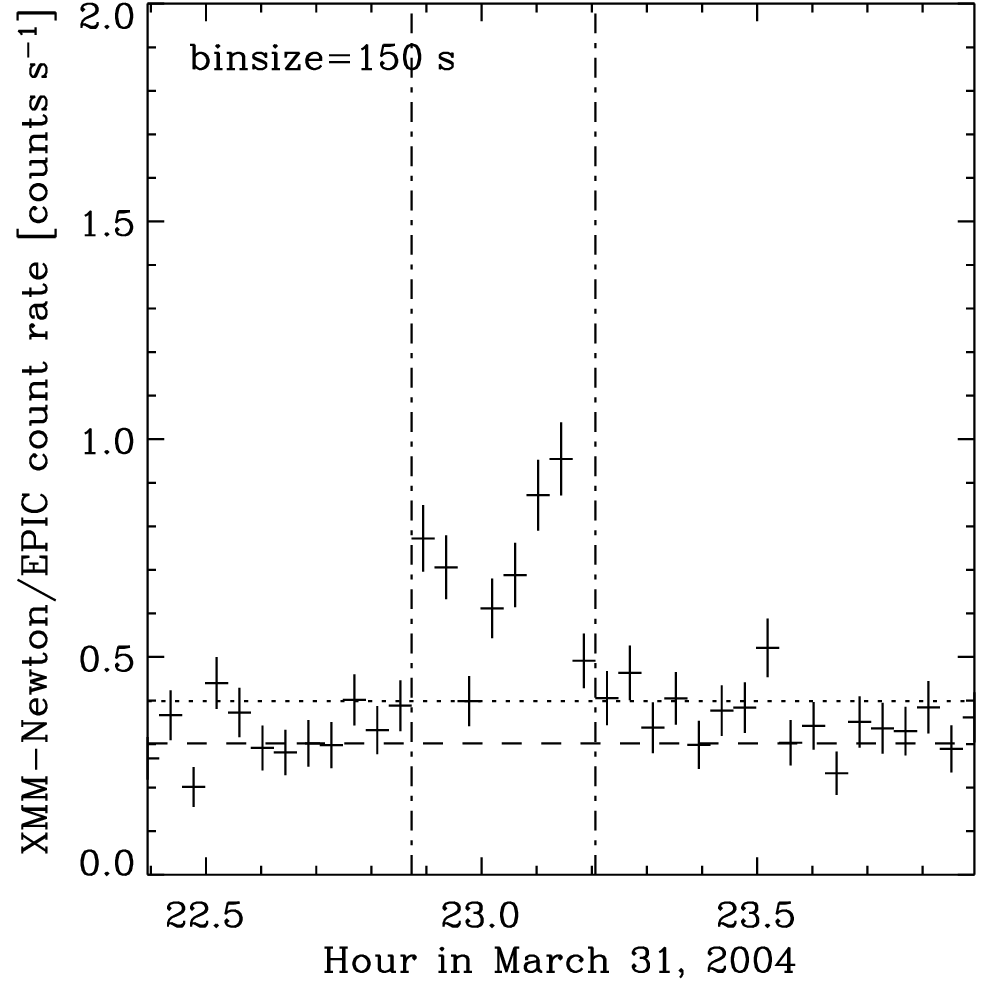}
\end{tabular}
\caption{Light curves of reprocessed X-ray anterior flares observed
  with {\sl XMM-Newton}. {\it Left}: October 2002 flare. {\it Right}: March
  2004 flare.}
\label{fig:zoom_anteriorflare}
\end{figure}

\begin{figure}[!ht]
\includegraphics[angle=-90,width=\columnwidth]{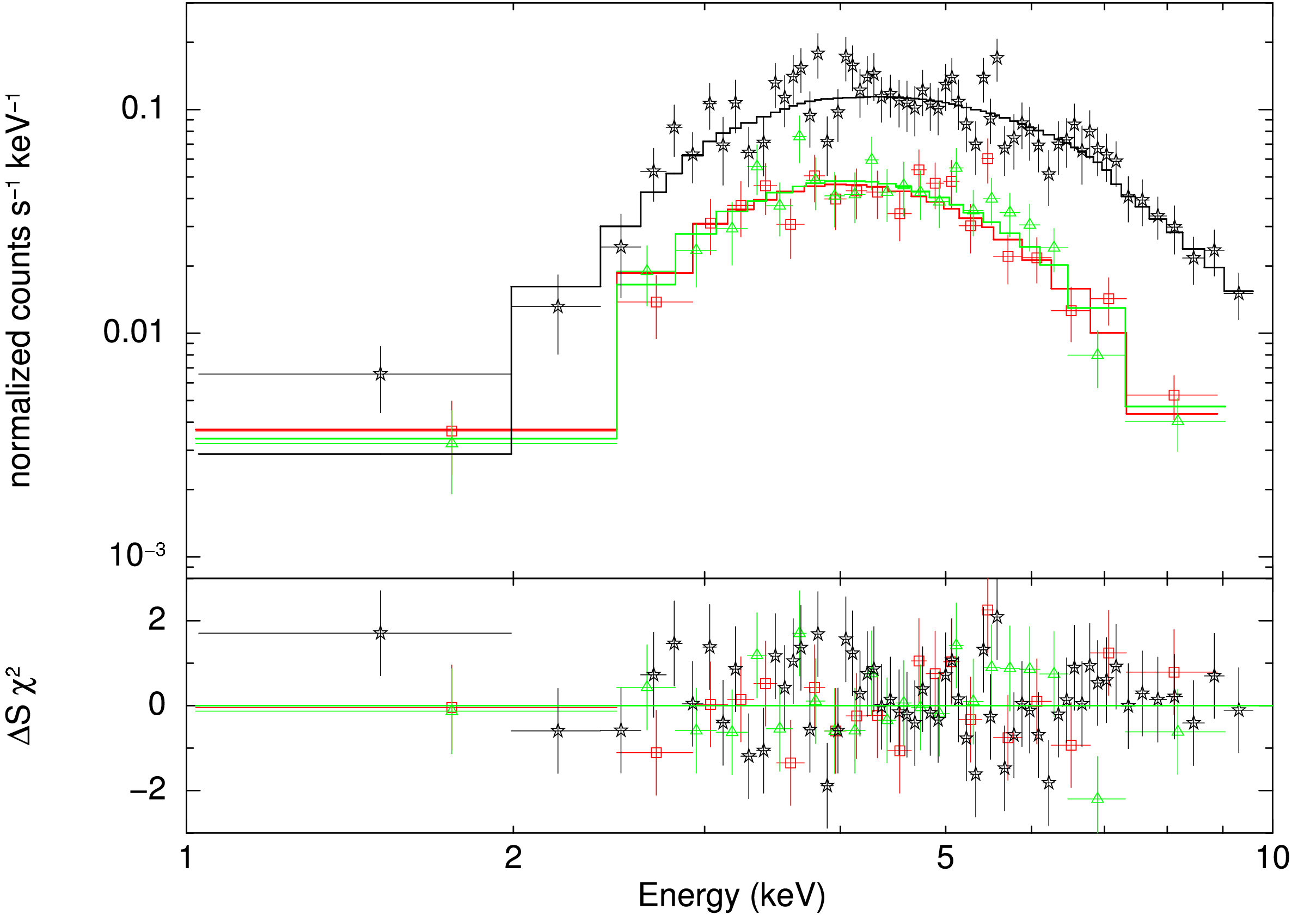}
\caption{{\sl XMM-Newton} EPIC spectra of the flare observed on
  October 2002. 
 The spectrum for the non-flaring period with the same extraction region
  is used as background. 
Squares, triangles and stars indicate MOS1 (red), MOS2 (green), and
  pn (black) data, respectively. 
The lines show the best-fit model using an absorbed power-law model
  taking into account dust
scattering (see Table~\ref{tab:fit} for best fit parameter values;
  {\sc pow} model). The
  fit residuals are shown in the bottom panels. 
}
\label{fig:plfit_oct02}
\end{figure}

\begin{figure}[!ht]
\includegraphics[angle=-90,width=\columnwidth]{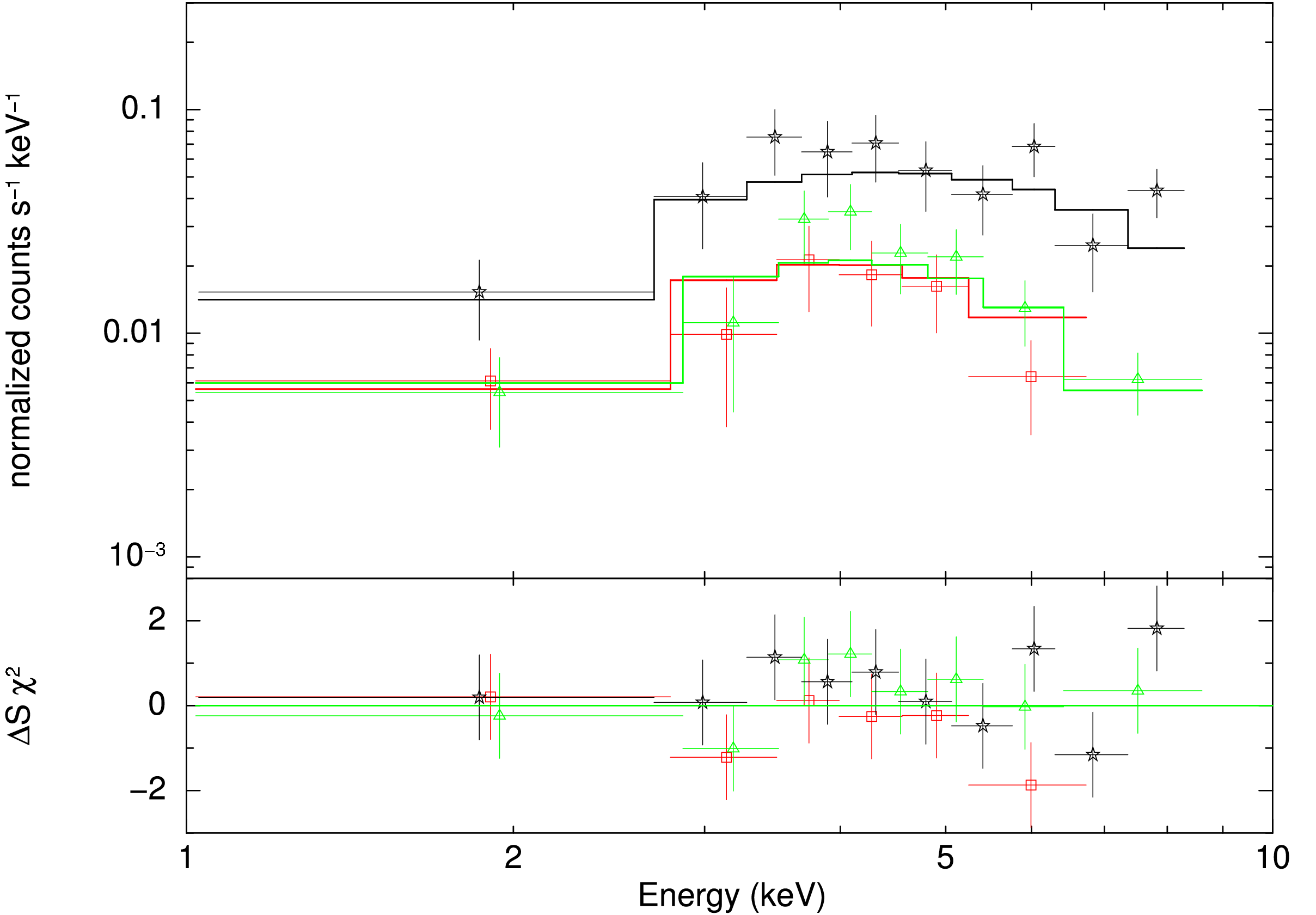}
\caption{Same as Fig.~\ref{fig:plfit_oct02} but for the flare observed on March 2004.}
\label{fig:plfit_march04}
\end{figure}

We report the reprocessed light curves
(Fig.~\ref{fig:zoom_anteriorflare}) and spectra
(Fig.~\ref{fig:plfit_oct02} and \ref{fig:plfit_march04}) of the X-ray flares observed on October 2002
and March 2004 that were reported previously by \citet{P03c} and
\citet{Belanger05}, respectively. Table~\ref{tab:flaretime_2} gives for each flare: the
start and end times, the duration, the total EPIC quiescent-subtracted
counts, and EPIC quiescent-subtracted count rate at the peak flare.

The X-ray flare observed on March 2004 was contaminated by a high
level of background proton flares which lead for pn to numerous
switches between science and counting mode, and therefore to a
significant loss of exposure time. Consequently, a bin time interval of
100\,s contains less than 50\% of (scientific mode) exposure, which is
not enough to obtain a good estimate of the count rate. We used a bin
time interval of 150\,s, which is sufficient to obtain a good estimate of the
count rate, when comparing with the MOS light curves. Obviously, this
loss of exposure on this moderate flare affected the quality of its pn
spectrum, and the resulting constraints derived on its physical parameters.

\begin{table*}[!ht]
\caption{Characteristics of the X-ray flares observed in October 2002
  and March 2004  (see Fig.~~\ref{fig:zoom_anteriorflare}). 
}
\label{tab:flaretime_2}
\begin{center}
\begin{tabular}{@{}ccccccrrrrr@{}}
\hline
\hline
\noalign {\smallskip}                       
\multicolumn{1}{c}{\small Flare} 
&\multicolumn{2}{c}{\small Start time$^\mathrm{(a)}$}
&\multicolumn{2}{c}{\small End time$^\mathrm{(a)}$}
&\multicolumn{1}{c}{\small Duration}
&\multicolumn{1}{c}{\small Total$^\mathrm{(b)}$}
&\multicolumn{1}{c}{\small Peak$^\mathrm{(c)}$} 
&\multicolumn{1}{c}{\small Det.$^\mathrm{(d)}$}
&\multicolumn{1}{c}{\small $L_{\rm 2-10~keV}^{\rm peak~(e)}$}
&\multicolumn{1}{c}{\small Ampl.$^\mathrm{(f)}$}\\
\noalign {\smallskip}                       
&\multicolumn{1}{c}{\small  hh:mm:ss}
& \multicolumn{1}{c}{\small s} 
&\multicolumn{1}{c}{\small hh:mm:ss}
&\multicolumn{1}{c}{\small s} 
& \multicolumn{1}{c}{\small s}
&\multicolumn{1}{c}{\small cts}
&\multicolumn{1}{c}{\small cts\,s$^{-1}$}
&\multicolumn{1}{c}{\small $\sigma$}
&\multicolumn{1}{c}{\small 10$^{34}$\,erg\,s$^{-1}$}\\
\noalign {\smallskip} 
\hline
\noalign {\smallskip} 
{\small 03/10/2002}& {\small 10:07:40 } & {\small  150026860} &     {\small 10:54:20} &{\small 150029660 }  & {\small
  2800}  & {\small 2158} & {\small 1.510} & {\small 29.6} & 35.9$^{+7.1}_{-5.2}$ & $^\mathrm{(g)}$149$^{+21}_{-16}$\\     
\noalign {\smallskip} 
{\small 31/03/2004} & {\small 22:52:24} & {\small 197160744}  &     {\small 23:12:24} & {\small  197161944} & {\small 1200} & {\small 447} & {\small 0.653}& {\small 12.8} & 11.0$^{+0.2}_{-0.2}$ & 46$^{+9}_{-15}$\\     
\noalign {\smallskip} 
\hline
\end{tabular}
\end{center}
\vspace*{-0.5cm}
\begin{list}{}{}
\item[$^{\mathrm{(a)}}$] Start and end times of the flare time
  interval defined as the period where the 
  EPIC light curve deviates from the non-flaring level at a confidence of
  95\%. See $\S$\ref{sec:lc} for details.  
\item[$^{\mathrm{(b)}}$] Total EPIC counts in the
  2--10\,keV energy band obtained during the flare interval
   after subtraction of the non-flaring level.
\item[$^{\mathrm{(c)}}$] EPIC count rate in the 2--10\,keV energy band
  at the flare peak after subtraction of the non-flaring level. 
\item[$^{\mathrm{(d)}}$] Detection level at the flare peak in $\sigma$.
\item[$^{\mathrm{(e)}}$] 2--10\,keV luminosity at the flare peak,
  assuming an absorbed power law model (taking account dust scatter
  effect), see parameter fits in Table~\ref{tab:fitCstat}. 
\item[$^{\mathrm{(f)}}$] Amplitude of the flare defined as the ratio of
the 2--10\,keV flare luminosity at the peak and the 2--10\,keV 
quiescent luminosity of Sgr\,A* observed with {\sl Chandra} 
(i.e., 2.4 $\times$ 10$^{33}$ erg\,s$^{-1}$, \citealt{Baganoff03}, and
this work, see $\S$\ref{sec:comparison}). 
\item[$^{\mathrm{(g)}}$] In \cite{P03c}, the amplitude was calculated
  using the 2--10\,keV quiescent luminosity of Sgr\,A* reported in
  \cite{Baganoff01}, i.e. 2.2 $\times$ 10$^{33}$ erg\,s$^{-1}$. Using this
  latter quiescent luminosity value, we would find for October 2002, an
  amplitude of 163$^{+23}_{-17}$.
\end{list}
\end{table*}

\end{appendix}

\end{document}